\documentclass[%
 reprint,
 amsmath,amssymb,
 aps,
]{revtex4-2}

\usepackage{graphicx}
\usepackage{dcolumn}
\usepackage{bm}
\usepackage{subcaption} 
\usepackage{lineno}

\begin{document}
\newcommand {\pt}{\ensuremath{p_{\text{T}}} }
\newcommand {\nch}{\ensuremath{N_{\text{ch}}} }
\newcommand {\nchfw}{\ensuremath{N_{\text{ch}}^{\text{FW}}} }
\newcommand {\nchue}{\ensuremath{N_{\text{ch}}^{\text{UE}}} }
\newcommand{\dphi}{\ensuremath{\Delta\phi} }
\newcommand{\deta}{\ensuremath{\Delta\eta} }
\newcommand {\vndelta}{\ensuremath{V_{\textrm{n}\Delta}} }
\newcommand {\vonedelta}{\ensuremath{V_{1\Delta}} }
\newcommand {\vtwodelta}{\ensuremath{V_{2\Delta}} }
\newcommand {\vthreedelta}{\ensuremath{V_{3\Delta}} }
\newcommand {\vndr}{\ensuremath{\frac{\vndelta}{\vndelta^{0-10} }} }
\newcommand {\vonedr}{\ensuremath{\frac{\vonedelta}{\vonedelta^{0-10} }} }
\newcommand {\vtwodr}{\ensuremath{\frac{\vtwodelta}{\vtwodelta^{0-10} }} }
\newcommand {\vthreedr}{\ensuremath{\frac{\vthreedelta}{\vthreedelta^{0-10} }} }


\title{Study of the multiplicity dependence of non-flow correlations in pp collisions at $\sqrt{s}=200$ GeV using PYTHIA}

\author{Milan Stojanovic}
\affiliation{Wayne State University}

\author{Joern Putschke}
\affiliation{Wayne State University}

\date{\today}

\begin{abstract}
The observation of a near-side enhancement in long-range correlations within high-multiplicity pA and pp collisions, commonly referred to as the ridge, has motivated significant efforts to quantify possible collective effects in small systems. 
Estimation of non-flow contributions in two-particle correlation measurements in small systems at the LHC typically relies on the assumption that jet-induced non-flow correlations are independent of event multiplicity. Consequently, this assumption allows the high-multiplicity non-flow to be estimated and accounted for by directly using low-multiplicity events.
In this paper, the validity of this assumption is studied examining the multiplicity dependence of non-flow contributions estimated via two-particle correlation functions in pp collisions at RHIC energies $\sqrt{s} = 200$~GeV using the PYTHIA event generator. It is observed that selecting any specific multiplicity range inherently 
biases jet production dynamics. These biases introduce a non-trivial multiplicity dependence of the non-flow contribution at RHIC energies. The dependence of jet fragmentation and $\pt$-differential cross sections on event multiplicity is studied using a variety of commonly used multiplicity estimators, including mid-rapidity charged-particle multiplicity, forward multiplicity, and underlying-event activity. The results indicate that non-flow contributions do not exhibit simple multiplicity scaling and challenge the applicability of standard subtraction procedures utilized at LHC energies at lower RHIC collision energies.
\end{abstract}

\maketitle

\section{Introduction}
 
Heavy-ion collisions at ultra-relativistic energies create temperatures and energy densities sufficient to form a new state of matter, commonly referred to as the quark–gluon plasma (QGP)~\cite{BRAHMS:2004adc,PHOBOS:2005,STAR:2005,PHENIX:2005}. In this deconfined phase, quarks and gluons are no longer bound within hadrons but instead they are quasi free, forming a strongly interacting, thermalized medium~\cite{Karsch:lqcd}. The key experimental signatures of this medium were discoveries of
the energy loss and modification of high-energy partons (jet quenching) combined with
azimuthal anisotropies in the momentum distribution of produced particles~\cite{Snellings:2011sz,Gyulassy:2003mc,Qin:2015srf}. These anisotropies arise from the asymmetry of the gradients, and are well described by relativistic hydrodynamic expansion of the system~\cite{Gale:2012rq}.

Surprisingly, similar azimuthal anisotropies have also been observed in high-multiplicity proton-proton (pp) and proton-nucleus (pA) collisions, systems traditionally considered too small to support the formation of a strongly interacting medium~\cite{CMS:2010ifv,CMS:2016fnw,ATLAS:2015hzw,ALICE:2012eyl,PHENIX:2014fnc,PHENIX:2015idk,STAR:2022pfn,STAR:2022pfn,STAR:2015kak}. Although the qualitative features of these correlations resemble those observed in heavy-ion collisions, their underlying physical mechanism is still 
being actively investigated in experiment and theory.
Possible explanations include both the existence of a medium-like state in small systems or alternatively momentum correlations present already at the earliest stages of the collision~\cite{Dusling:2015gta,Nagle:2018nvi,Grosse-Oetringhaus:2024bwr}. Although some initial-state models have shown that initial state momentum correlations cannot create correlations across large rapidity intervals~\cite{Schenke:2022mjv}, final-state suppression effects have not yet been observed in small systems (smaller than O+O collisions)~\cite{PHENIX:2003qdw,ALICE:2014nqx,CMS:2016xef}, and theoretical calculations indicate that energy loss cannot explain the observed collective flow in pA collisions~\cite{Bert:2026uxa}.
One of the main obstacles towards a better understanding of non-flow is the absence of a universally accepted method for quantifying it. In particular, the applicability of techniques developed for large nucleus-nucleus (AA) collisions is limited due to the significant contribution of few-particle correlations, for example jet production, commonly referred to as non-flow, which can strongly affect the measurement in small collision systems.

A widely used approach to study azimuthal anisotropies is based on two-particle correlation functions, constructed as a function of the relative azimuthal angle \dphi (and often pseudorapidity separation \deta)~\cite{Ollitrault:1993,Voloshin:1994mz,Poskanzer:1998yz}. When particle production is dominated by global, collective dynamics, particles are correlated with a common symmetry plane, and consequently, with each other. This results in a characteristic modulation of the correlation function, with pronounced structures at $\dphi = 0$ and, by symmetry, at $\dphi = \pi$. Contributions from few-particle correlations, those originating from resonance decays and jet fragmentation, are typically localized in a narrow ($\deta,\dphi$) window around (0,0) (near-side) and, therefore, can be largely suppressed by requiring a pseudorapidity separation. The modulation can be easily quantified with coefficients of Fourier expansion: 
\begin{equation}
\label{eq:FourierDec}
\frac{ \text{d}N^{pairs}}{\text{d}\dphi} \propto  1 + \sum\limits_{n=1}^{\infty}2\vndelta\cos[n(\dphi)].
\end{equation}

Jets are, however, typically produced in pairs with approximately back-to-back kinematics. In heavy-ion collisions those correlations are significantly reduced 
since the contribution of jet fragments to the total particle yield is relatively small. Additionally, the jets are suppressed 
due to strong interactions of partons with the medium. In small systems, by contrast, there is no significant jet quenching, and the relative contribution of jet fragments to the total event multiplicity is substantially larger. As a result, back-to-back jet pairs remain a dominant source of correlations, producing a pronounced structure at $\dphi \approx \pi$ (away-side), which remains largely independent on pseudorapidity. Therefore, these correlations cannot be efficiently removed by a \deta selection and constitute a significant non-flow contribution in small systems.


To separate possible collective from jet-induced effects in two-particle correlation measurements, several non-flow subtraction methods have been developed~\cite{CMS:2016fnw,ATLAS:2015hzw}. Most commonly, these approaches rely on the assumption that non-flow correlations scale universally with event multiplicity and that their contribution can be removed by contrasting correlation functions at different multiplicities, as implemented in low-multiplicity subtraction and template fit approaches. However, to utilized these methods, further assumptions on the evolution of collective effects with multiplicity are needed. For instance, the low-multiplicity subtraction method assumes that collectivity is entirely absent in low-multiplicity events, whereas the template fit method assumes that the first term in Eq.~\ref{eq:FourierDec} is dominated strictly by non-flow.
That leads to the situation where existing measurements reported by CMS and ATLAS exhibit notable differences in the extracted flow coefficients in small systems~\cite{ATLAS:2017rtr}. 
The discrepancy reflects the sensitivity of the results on the underlying subtraction method and its physics assumptions with respect to collective phenomena. 

Alternative approaches for suppressing non-flow contributions have also been developed. In particular, multi-particle cumulant techniques reduce the contribution from few-particle correlations by measuring correlations among 4, 6, 8, or even 10 particles~\cite{Voloshin:1994mz,Bilandzic:2013kga}. More recently, sub-event cumulant methods, with the additional condition that particles fall into disjoint pseudorapidity regions, have demonstrated particularly strong suppression of non-flow effects~\cite{Jia:2017hbm}. However, these methods generally require substantially larger statistics, higher multiplicities, and sufficiently strong collective signals. In addition, analyses in pA collisions have explored alternative strategies based on vetoing events containing reconstructed jets, thereby directly reducing the contribution of jet-induced correlations to the measured observables~\cite{CMS:2022bmk}.

Despite the development of a variety of techniques for suppressing these contributions, the underlying interplay between jet production and event activity is still not fully understood. In fact, recent model studies indicate that at RHIC energies the subtraction procedures do not perform reliably at moderate multiplicities~\cite{Lim:2019cys}. 
Largely because $\dphi$ correlations evolve dramatically across different multiplicity intervals, hence the resulting flow signals become highly sensitive to the choice of the low-multiplicity baseline.
This observation motivates a more detailed investigation of the potential biases introduced by event multiplicity selection, in particular its impact on jet production and properties. Since jets constitute a dominant source of non-flow correlations in small systems, any bias of their kinematic distributions or fragmentation patterns with multiplicity can directly affect the non-flow corrected azimuthal anisotropies. 
Additionally, selecting rare high-activity events can preferentially sample particular correlations between high-$\pt$ partons and the soft particles used for multiplicity estimation. As a consequence, the fragmentation of hard partons and the resulting jet-induced correlations may become directly coupled to the event activity definition itself. Understanding the origin and strength of these hard-soft correlations is therefore essential for assessing the validity of non-flow subtraction methods and for interpreting measurements of collective behavior in small collision systems.

In this paper, we investigate whether non-flow subtraction methods based on two-particle correlations remain applicable at RHIC energies and qauntify the bias on non-flow estimations via commonly used multiplicity selections. For this purpose, pp collisions at $\sqrt{s}=200$~GeV are simulated using the PYTHIA event generator. The second section describes the simulation setup and jet reconstruction procedure. The third section introduces the different event activity estimators used in the analysis. The fourth section presents the construction of the two-particle correlation functions and investigates the applicability of standard subtraction approaches. The following section focuses on the effect of multiplicity selection on jet production and fragmentation observables. Finally, the discussion section summarizes the findings and considers further implications of these results on understanding the particle creation mechanisms in small systems.

\section{Monte Carlo Simulation and Jet Reconstruction}
\label{sec:MCandJets}
 
\subsection{Event Generation}
\label{subsec:MC}

To provide a baseline for particle production, Monte Carlo simulations were performed using the \textsc{PYTHIA} 8.314 event generator~\cite{Bierlich:2022pfr}. This framework enables a controlled study of particle correlations arising from QCD processes in the absence of collective effects.

Minimum-bias $p+p$ collisions were simulated at $\sqrt{s}=200$~GeV (corresponding to typical RHIC energies) using the \texttt{SoftQCD:inelastic = on} setting in \textsc{PYTHIA}~8. To evaluate the impact of QCD processes, such as multiparton interactions (MPI), initial- and final-state radiation, and color reconnection, two distinct tunes were tested: the default Monash~2013~\cite{Skands:2014pea} tune (optimized primarily for LHC data) and the Detroit~\cite{Aguilar:2021sfa} tune (customized to describe RHIC $p+p$ observables at $\sqrt{s}=200$~GeV), generating a total sample of one billion events for each configuration.

\subsection{Jet Reconstruction}
\label{subsec:Jets}

Jets were reconstructed using the FastJet 3.5.1 package with the anti-$k_{\text{T}}$ clustering algorithm~\cite{Cacciari:2011ma,Cacciari:2005hq,Cacciari:2008gp}. The resolution parameter was set to $R = 0.4$, and the recombination scheme followed the standard $E_{\text{T}}$-scheme.

Input particles for jet reconstruction were taken from final-state charged particles with $\pt > 0.3$~GeV and $|\eta| < 1.5$, using their true four-momenta. Jet reconstruction was performed within the kinematic acceptance $|\eta| < 1.5$.

\section{Event activity estimator}

\begin{figure*}[htbp]
     \centering
         \includegraphics[width=0.7\textwidth]{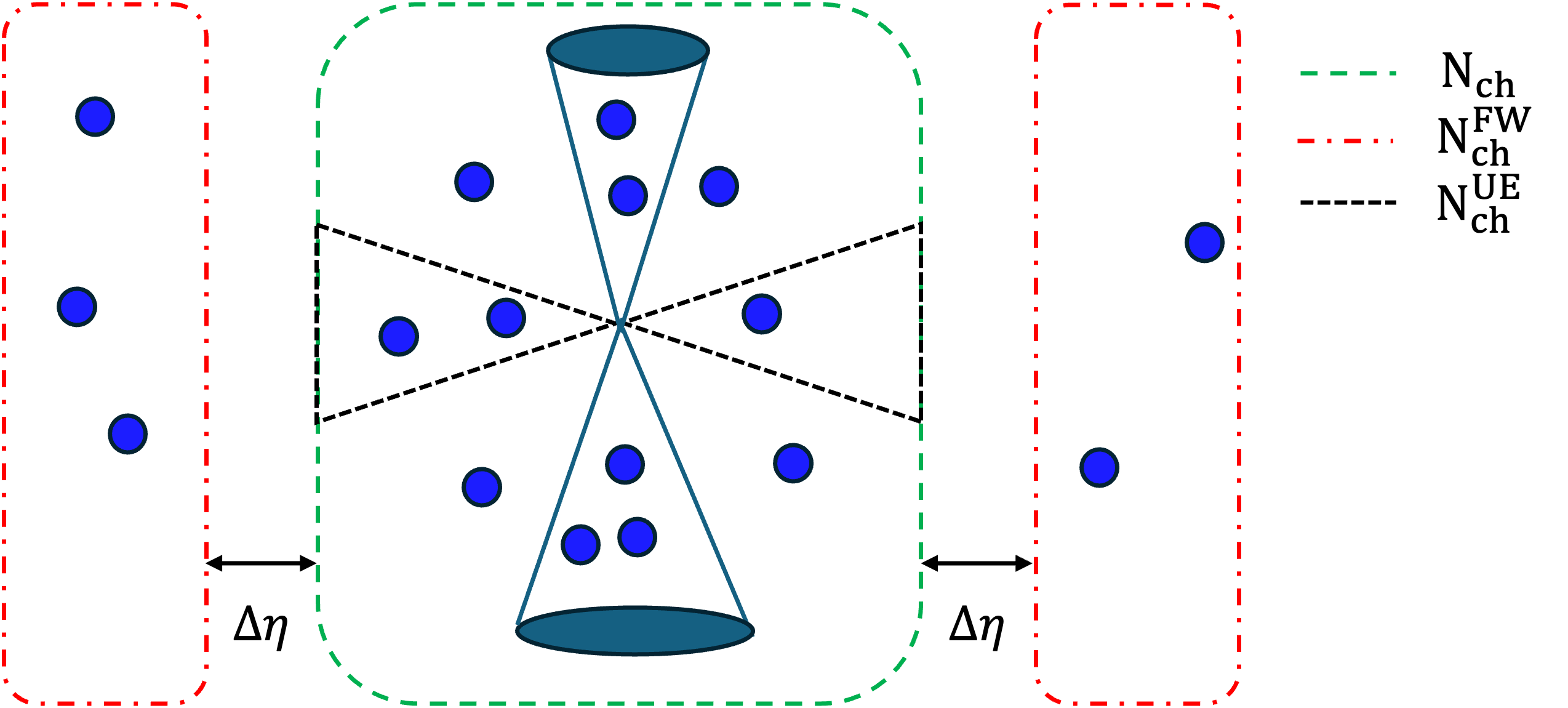}
        \caption{Illustration of the three event multiplicity estimators considered in this work, shown for the same event. The captured regions indicate the phase-space acceptance used for each multiplicity definition in the same event: nominal midrapidity multiplicity (green), forward multiplicity (red), and underlying-event multiplicity (black).}
     \label{fig:cartoonAll}
\end{figure*}

To investigate the interplay between event activity selection and jet production, three different event multiplicity estimators are considered in this work: charged-particle multiplicity at midrapidity, forward rapidity, and multiplicity measured in the underlying-event region, as illustrated in Fig.~\ref{fig:cartoonAll}. These estimators are designed to probe the sensitivity of (anti-)correlation between the multiplicity definition and the reconstructed jet population, thereby allowing a systematic study of multiplicity-induced biases on non-flow correlation estimates. 

\textbf{Midrapidity multiplicity estimator}

The default event activity estimator (\nch) is defined as the number of charged particles with $0.3 < \pt < 3$~GeV and pseudorapidity $|\eta| < 1.5$. This phase space corresponds to the region used for the two-particle correlation analysis and jet reconstruction. Such a choice is commonly employed in small-system analyses due to its direct connection to the particle density in the region where collective signatures are measured. 

Using the same phase space for multiplicity estimation and correlation measurements can potentially introduce bias. Particles originating from jet fragmentation contribute simultaneously to the reconstructed multiplicity and to the measured correlation structure. As a consequence, selecting events in different multiplicity intervals could sample events with different jet topologies and fragmentation patterns, potentially modifying the observed non-flow contributions.

\textbf{Forward multiplicity estimator}

To reduce the direct overlap between the multiplicity definition and the jet reconstruction region forward (\nchfw) event activity estimator is constructed using charged particles with $0.3 < \pt < 3$~GeV in the forward pseudorapidity interval $3.5 < |\eta| < 5$. In this configuration, the multiplicity estimator is separated from the midrapidity region where jets and two-particle correlations are measured. 

\begin{figure}[htbp]
     \centering
     \begin{subfigure}[b]{0.45\textwidth}
         \centering
         \includegraphics[width=\textwidth]{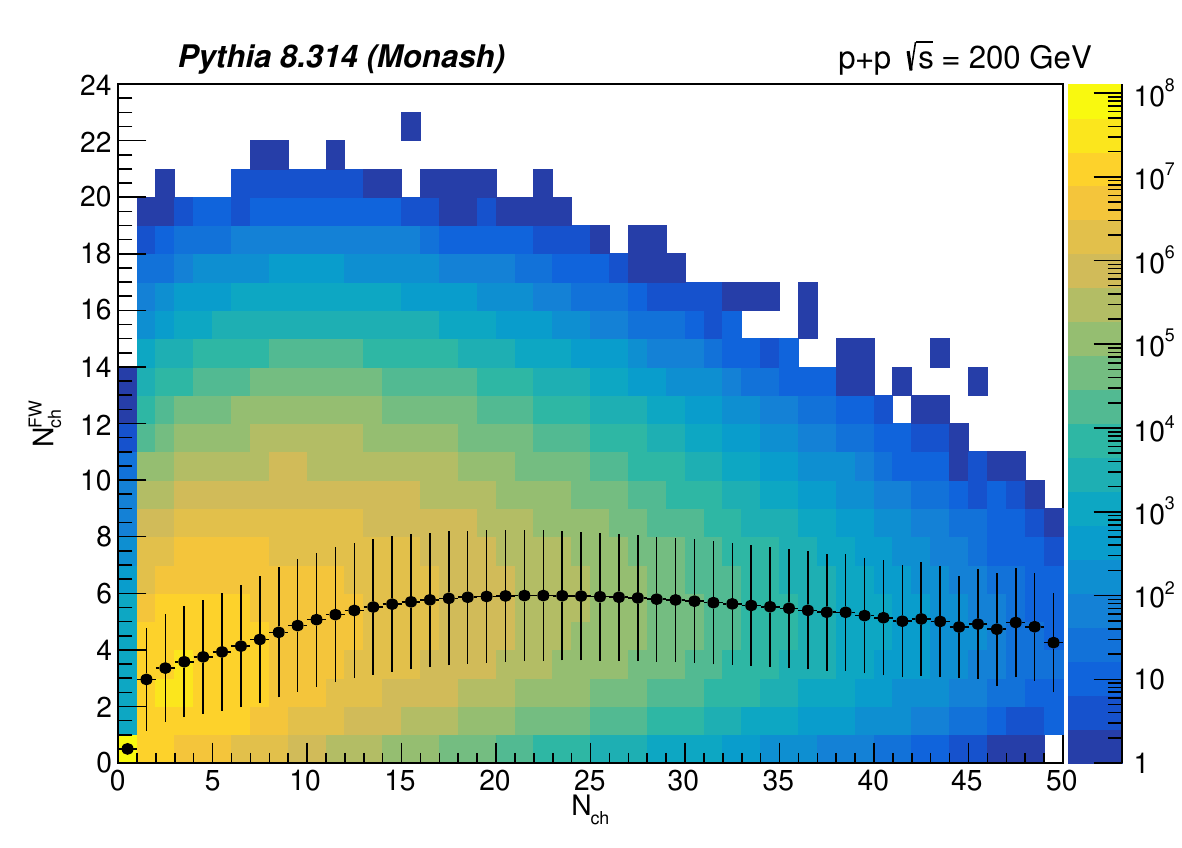}
         \caption{\label{fig:MultCorr1}}
        
     \end{subfigure}
     \hfill
     \begin{subfigure}[b]{0.45\textwidth}
         \centering
         \includegraphics[width=\textwidth]{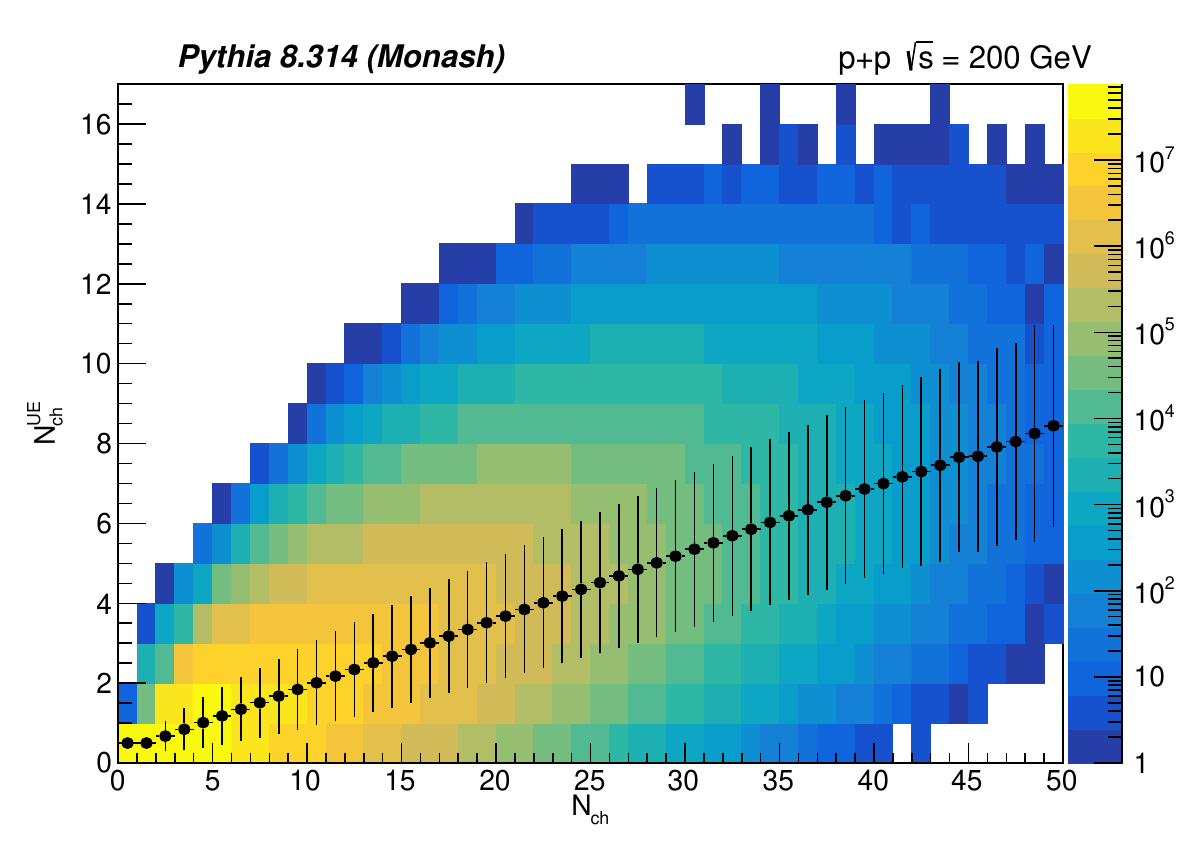}
         \caption{\label{fig:MultCorr2}}    
     \end{subfigure}

     \caption{Correlation between event multiplicities estimated at mid- and forward rapidity (a), and at midrapidity and in the underlying-event region (b). The black points and error bars represent the mean value and RMS of the multiplicity measured with the alternative event activity estimator for a given nominal multiplicity interval, respectively.}
     \label{fig:MultCorr}
\end{figure}

This approach suppresses direct autocorrelations associated with jet production entering the multiplicity definition~\cite{ALICE:2024dcr}. Although event activity measured at forward rapidity at the LHC exhibits strong correlations with midrapidity multiplicity, our study indicates that at RHIC energies the correlation between the number of particles emitted in these two widely separated kinematic regions is significantly weaker. This conclusion consistent with previous observations by the STAR Collaboration~\cite{STAR:2024nwm}. As shown in Fig.~\ref{fig:MultCorr1}, the midrapidity and forward multiplicity estimators exhibit positive correlations only at relatively low multiplicities, roughly $\nch \lesssim 15$, with a weak anticorrelation developing between forward and midrapidity particle production at higher \nch.


\textbf{Underlying-event multiplicity estimator}

A third event activity estimator is defined using particles in the underlying-event region relative to the reconstructed leading jet axis (\nchue). In this approach, charged particles, with the same \pt interval as previous two estimators: $0.3 < \pt < 3$~GeV but separated from the leading jet axis by $\dphi > 1$ are used to define the event multiplicity. By construction, this estimator suppresses the direct contribution of particles associated with the reconstructed jet cone. 
Compared to the forward-region estimator, the underlying-event multiplicity exhibits stronger correlations with the midrapidity multiplicity, as shown in Fig.~\ref{fig:MultCorr2}. However, despite the approximately linear trend across the full multiplicity range, the distribution develops significantly larger fluctuations in high-multiplicity events.

\begin{figure}[htbp]
     \centering
         \includegraphics[width=0.48\textwidth]{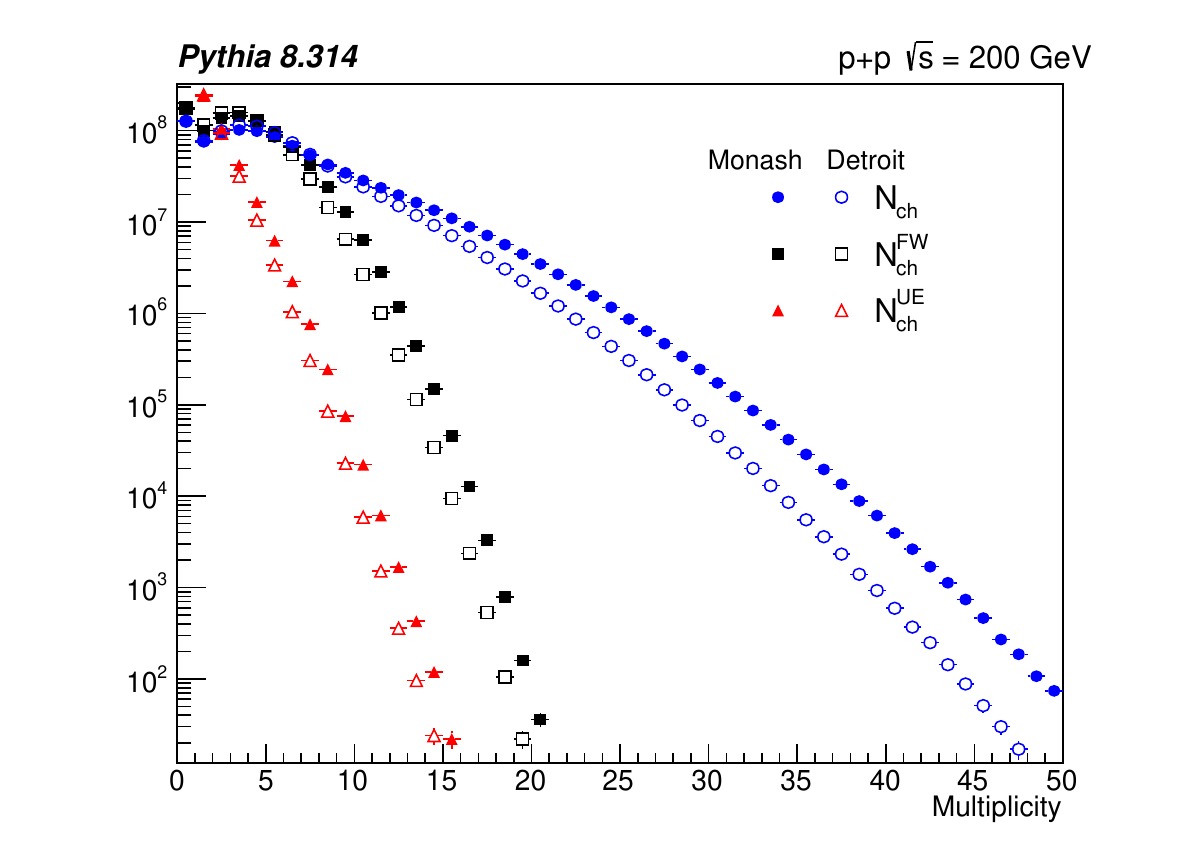}
             
     \caption{Charged-particle multiplicity distributions for the midrapidity, forward-rapidity, and underlying-event estimators, obtained with the Monash and Detroit PYTHIA tunes.}
     \label{fig:MultAll}
\end{figure}

The multiplicity distributions from all three estimators are shown in Fig.~\ref{fig:MultAll}. 
The multiplicity distributions from the Monash and Detroit tunes are very similar for all three event activity estimators,although the Detroit tune exhibits a slightly narrower distribution for all estimators.
This indicate that the overall event activity is largely insensitive to the choice of tune.

\section{Two-particle correlations analysis}

The non-flow contribution in generated events is investigated using well established methods based on two-particle azimuthal correlation functions~\cite{Voloshin:1994mz,Poskanzer:1998yz,Ollitrault:2009ie}. These distributions are constructed from the difference in azimuthal angle, $\dphi$, between all pairs of particles satisfying the selection criteria. The correlation function is expressed in terms of the per-particle associated yield as a function of $\dphi$.

The particles are required to be charged, with transverse momentum in the range $0.3 < \pt < 3$~GeV and pseudorapidity $|\eta| < 1.5$. The particle selection used for constructing the correlation functions is identical to that employed in the nominal event multiplicity definition, ensuring internal consistency of the analysis and following common practice in the literature. A pseudorapidity gap of $|\deta| > 2$ is imposed to suppress correlations among particles originating from the same jet (short-range correlations).

The azimuthal anisotropy coefficients, $\vndelta$, are extracted by performing a Fourier decomposition of the correlation function, including harmonics up to order $\text{n}=5$. 

\begin{figure}
 \includegraphics[width=0.48\textwidth]{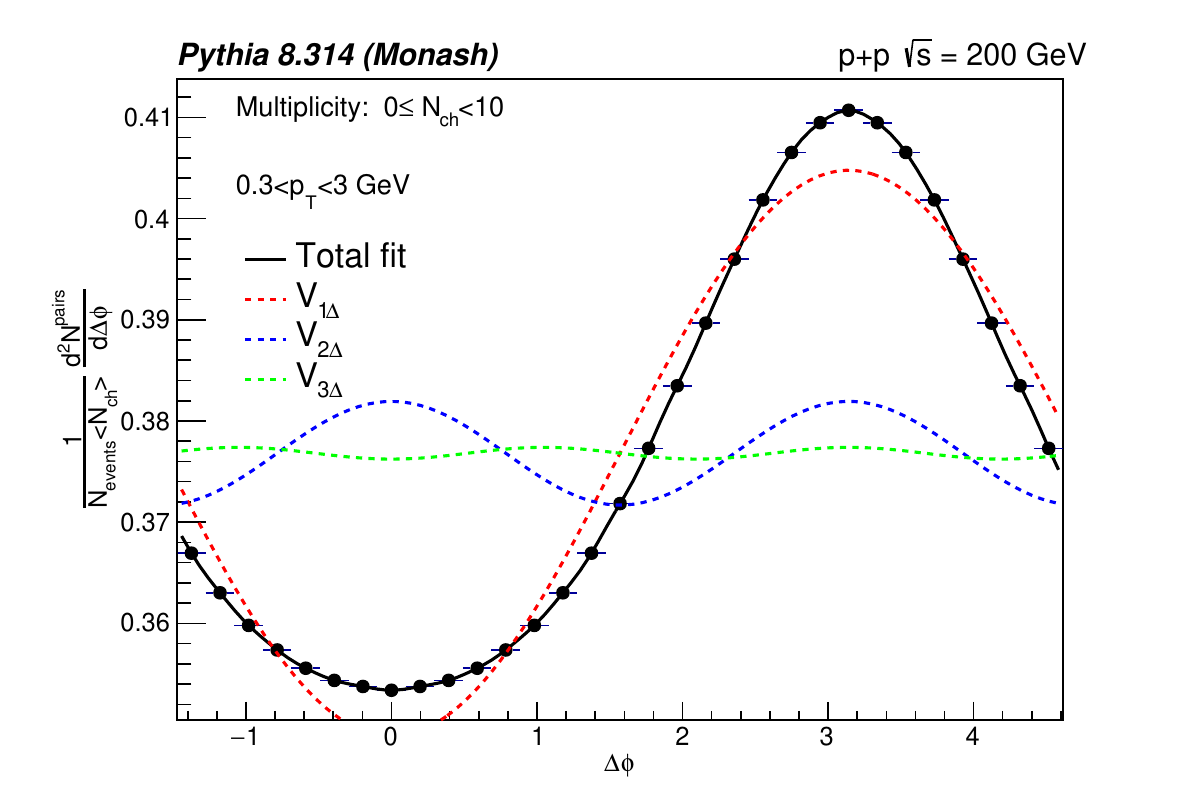}
 \caption{Two-particle azimuthal correlation function as a function of $\dphi$, for $0\leq\nch<10$. The lines represent the results of Fourier fit.\label{fig:Fourier}}
 \end{figure}

Figure~\ref{fig:Fourier} shows the resulting two-particle correlation distribution together with its Fourier decomposition, for the multiplicity class $0\leq\nch<10$
illustrating the contributions of the different harmonic components. 

The extracted harmonic coefficients for $\text{n} = 1, 2,$ and $3$ as a function of event multiplicity are shown in Table~\ref{tab:vn_results}. A clear decrease in the magnitude of all coefficients is observed with increasing multiplicity. This correlation strength dilution with increasing particle multiplicity is consistent with trends previously reported in LHC measurements~\cite{CMS:2016fnw} and studies based on PYTHIA simulations~\cite{Lim:2019cys}.

\begin{figure}
\includegraphics[width=0.48\textwidth]{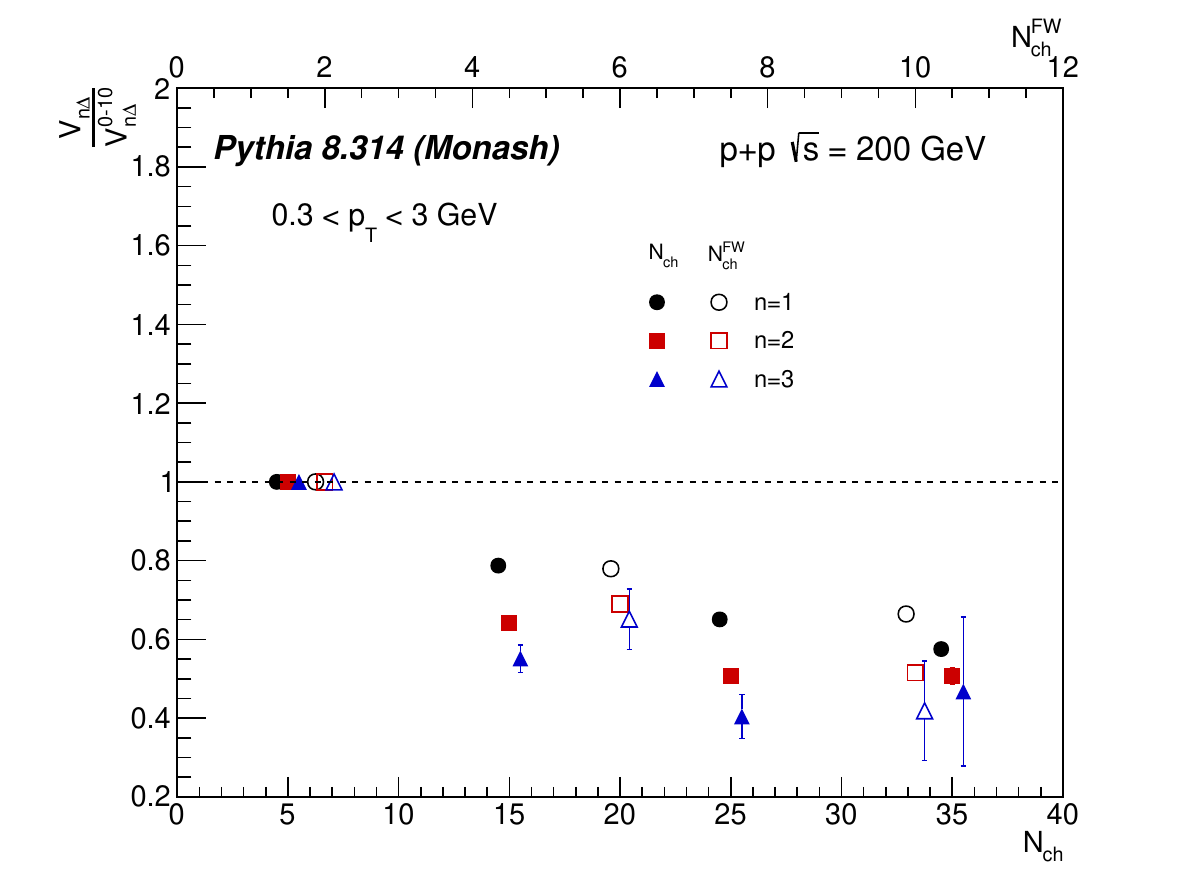}
 \caption{Azimuthal anisotropy coefficients \vonedelta, \vtwodelta, and \vthreedelta, normalized to their value in the lowest multiplicity bin, as a function of event multiplicity estimated at mid and  forward rapidity. The coefficients are extracted from a Fourier decomposition of two-particle correlation functions with a pseudorapidity gap of $|\deta| > 2$. \label{fig:double_ratio}}
 \end{figure}

However, the observed reduction is not uniform across the different $\vndelta$ coefficients, indicating a nontrivial non-flow dependence on multiplicity. To further quantify this effect, Fig.~\ref{fig:double_ratio} shows the multiplicity dependence of $\vndelta$ normalized to the value measured in the lowest multiplicity interval, $0 \leq \nch < 10$. This representation highlights the relative evolution of the different harmonic components and emphasizes deviations from a common scaling behavior.

Most non-flow subtraction methods rely on the assumption that non-flow contributions scale universally with event multiplicity~\cite{CMS:2016fnw,ATLAS:2015hzw}. In this framework, the total measured \vndelta is a combination of the flow $\vndelta^{\mathrm{flow}}$ and non-flow $\vndelta^{\mathrm{nonflow}}$ components, and the non-flow component of the harmonic coefficients in high-multiplicity (HM) events is assumed to be proportional to that in low-multiplicity (LM) events, i.e., $\vndelta^{\mathrm{nonflow,HM}} = c \cdot \vndelta^{\mathrm{nonflow,LM}}$, where the scaling factor $c$ is taken to be independent of multiplicity. 
However, the results presented in this work indicate that such a simple scaling does not hold. The multiplicity dependence of the $\vndelta$ coefficients exhibits a non-uniform behavior across different harmonics, inconsistent with a single, scaling factor, independent of harmonic order. This observation suggests that one of the key assumptions of non-flow subtraction methods breaks down in pp collisions at RHIC energies. Given that non-flow contributions are largely driven by back-to-back jet correlations, it is plausible that their structure and relative contribution evolve with event multiplicity, rather than following a simple scaling. That scenario is explored in more detail in this paper.

\begin{table}[htbp]
\centering
\scriptsize
\caption{Azimuthal anisotropy coefficients \vonedelta, \vtwodelta, and \vthreedelta for different multiplicity ranges.}
\label{tab:vn_results}
\setlength{\tabcolsep}{1pt}
\begin{tabular}{l|c|c|c}
\hline
\hline
\nch & \vonedelta & \vtwodelta & \vthreedelta \\
\hline
0-10 & $-0.03710 \pm 0.00003 $ & $0.00678 \pm 0.00003 $ & $-0.00077 \pm 0.00003 $ \\
10-20 & $-0.02921 \pm 0.00002 $ & $0.00435 \pm 0.00002 $ & $-0.00042 \pm 0.00002 $ \\
20-30 & $-0.024139 \pm 0.00004 $ & $0.003440 \pm 0.00004 $ & $-0.000310 \pm 0.00004 $ \\
30-40 & $-0.021343 \pm 0.00014 $ & $0.003437 \pm 0.00014 $ & $-0.00036 \pm 0.00014 $ \\
\hline
\hline
\end{tabular}
\end{table}

\begin{table}[htbp]
\centering
\scriptsize
\caption{Azimuthal anisotropy coefficients \vonedelta, \vtwodelta, and \vthreedelta for different \nchfw ranges.}
\label{tab:vnfw_results}
\setlength{\tabcolsep}{0.3pt}
\begin{tabular}{l|c|c|c}
\hline
\hline
\nchfw & \vonedelta & \vtwodelta & \vthreedelta \\
\hline
0-4 & $-0.036108 \pm 0.000047 $ & $0.005405 \pm 0.000047 $ & $-0.000617 \pm 0.000047 $ \\
4-8 & $-0.028138 \pm 0.000036 $ & $0.0.003727 \pm 0.000036 $ & $-0.000402 \pm 0.000036 $ \\
8-12 & $-0.02399 \pm 0.00007 $ & $0.00278 \pm 0.00007 $ & $-0.00025 \pm 0.00007 $ \\
\hline
\hline
\end{tabular}
\end{table}

The variation of the flow coefficients with \nchfw is found to be weaker than that observed with \nch. This behavior is expected due to the relatively weak correlation between \nch and \nchfw, given that the flow coefficients are measured in midrapidity in both cases.
Nevertheless, the Fourier coefficients do not exhibit a simple scaling with multiplicity, consistent with the behavior observed when the multiplicity and two-particle correlation functions are constructed within the same phase space. 
This behavior suggests that the observed correlation between back-to-back jet production and event multiplicity cannot be explained solely as a trivial consequence of overlapping phase-space selection.

\begin{figure}
\includegraphics[width=0.48\textwidth]{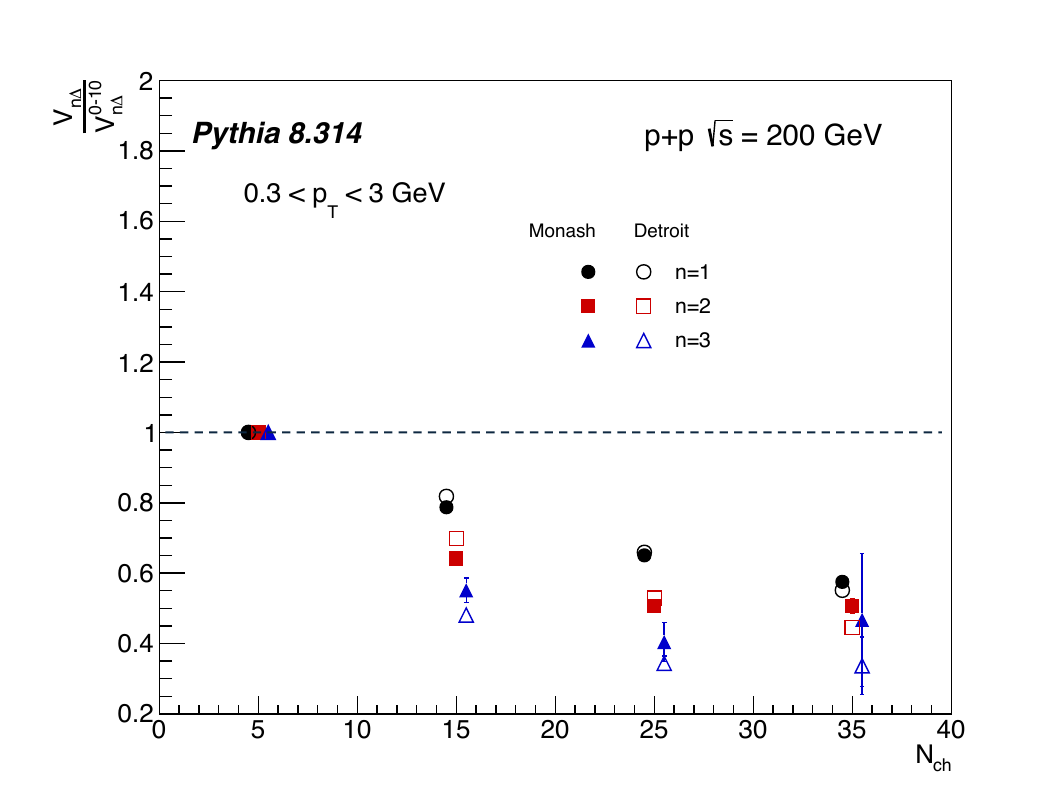}
 \caption{Azimuthal anisotropy coefficients \vonedelta, \vtwodelta, and \vthreedelta, normalized to their value in the lowest multiplicity bin, as a function of event multiplicity for Monash (closed symbols) and Detroit (open symbols) tunes. \label{fig:double_ratio_vsdetroit}}
 \end{figure}

To evaluate the sensitivity of these results to specific model parameterizations, the default Monash tune results are compared to the Detroit tune, the latter being specifically optimized to reproduce soft particle production at RHIC energies~\cite{Aguilar:2021sfa}. As shown in Fig.~\ref{fig:double_ratio_vsdetroit}, while minor quantitative differences emerge, the qualitatively observed behavior remains independent of the choice of PYTHIA tune, demonstrating the robustness of the extracted trends.

\subsection{Role of Particle Production Mechanisms and MPI}

Although the multiplicity-selection bias causing the non-scaling behavior of Fourier harmonics is expected to originate from particle production mechanisms, a similar trend persists across different PYTHIA tunes. 
To better understand how the interplay between hard and soft particle production mechanisms influences the evolution of non-flow correlations, it is essential to understand how those correlations depend on the relative contributions originating from hard jet fragmentation versus soft MPI to the total event multiplicity. 
In this study, the balance is controlled by changing the infrared regularization scale, $p_{\mathrm{T0}}$, which screens the divergence of the perturbative QCD cross section and directly regulates MPI activity.

\begin{figure}
    \centering
    \includegraphics[width=0.95\linewidth]{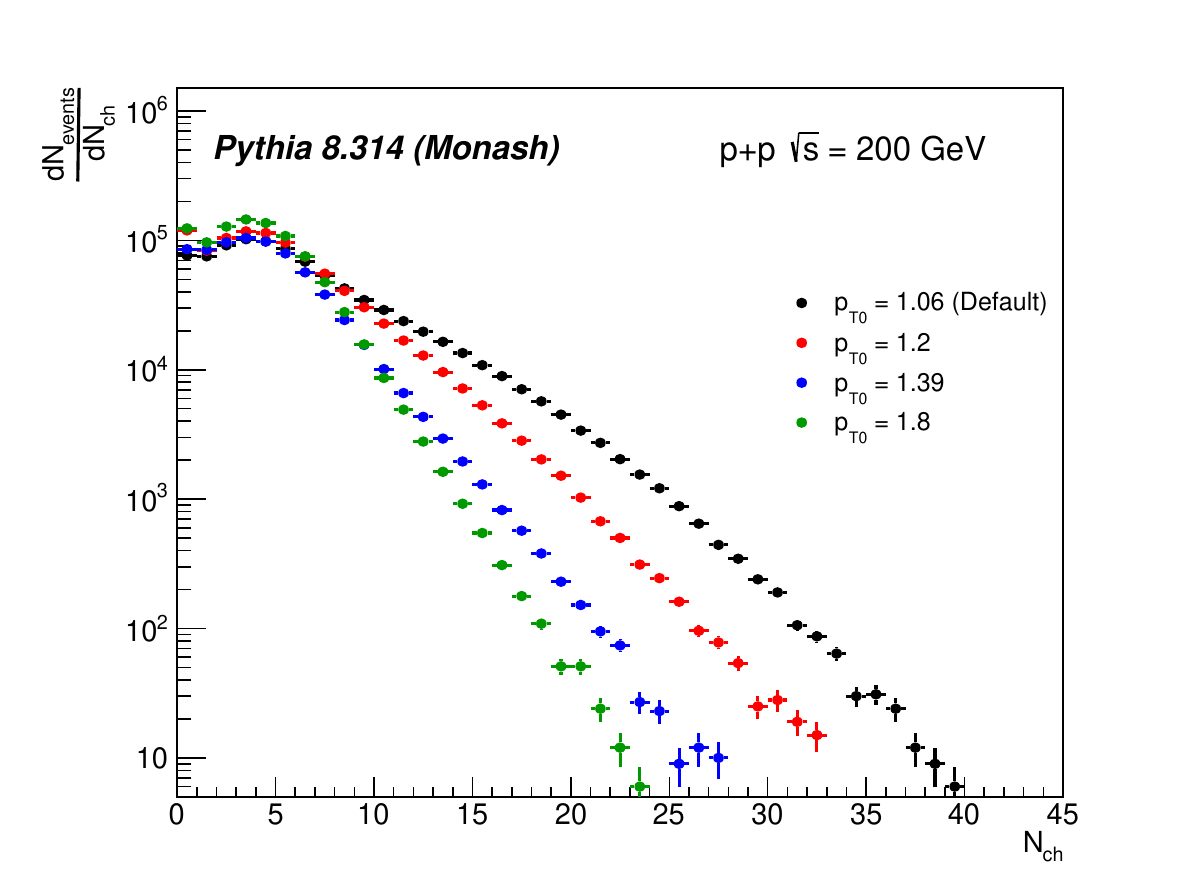}
    \includegraphics[width=0.95\linewidth]{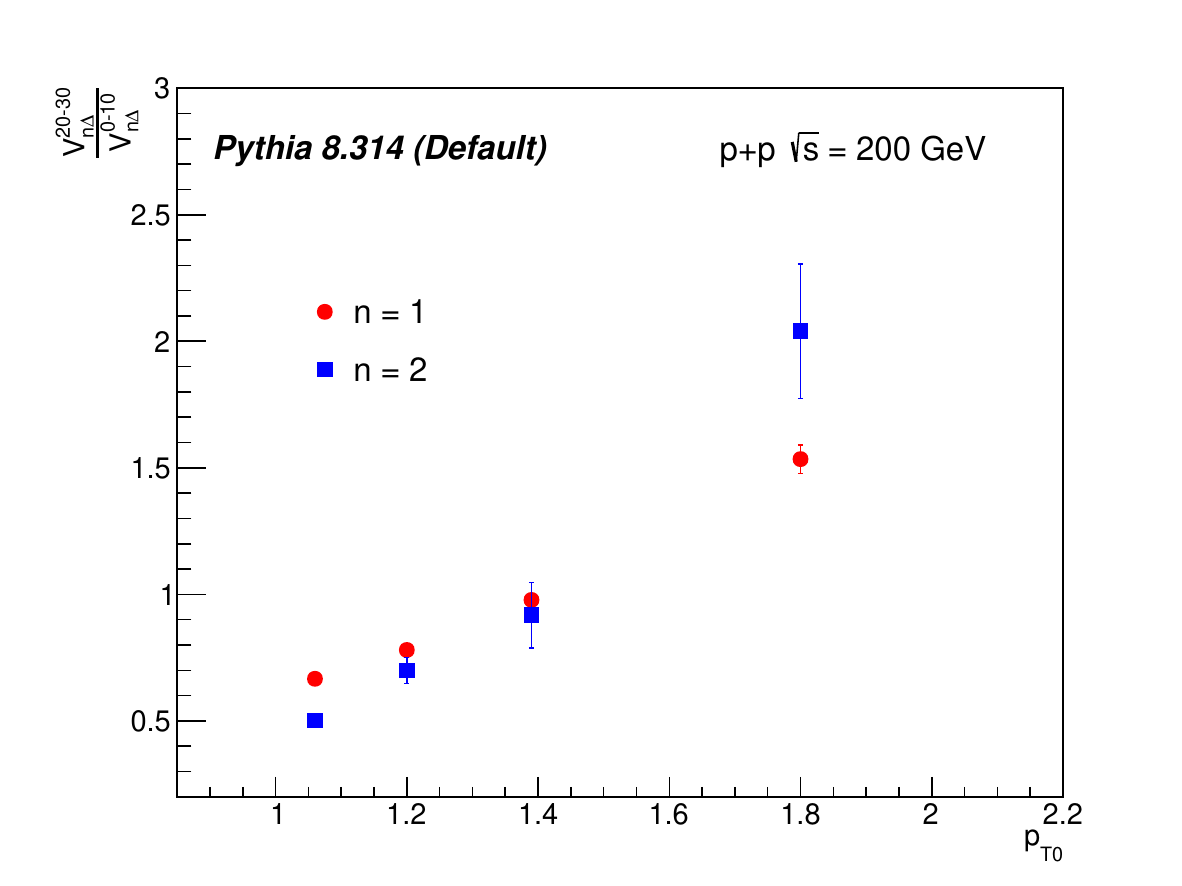}
    \caption{Impact of the infrared regularization scale $p_{\mathrm{T0}}$ on multiplicity distribution (top panel) and non-flow scaling (bottom panel).}
\label{fig:pt0_comparison}
\end{figure}

The top panel of Fig.~\ref{fig:pt0_comparison} illustrates the significant impact of $p_{\mathrm{T0}}$ on the overall event activity, where a larger regularization scale shifts the multiplicity distribution toward lower values, and severely truncates the high-multiplicity tail.

This modification of the hard-soft correlation dynamics has significant consequences for the evolution of the non-flow correlations. As shown in the bottom panel of Fig.~\ref{fig:pt0_comparison}, varying $p_{\mathrm{T0}}$  alters how the azimuthal anisotropy coefficients scale with event multiplicity, $N_{\mathrm{ch}}$. The ratio of anisotropy coefficients between high and low multiplicity intervals, 
$\frac{\vndelta^{20-30}}{\vndelta^{0-10}}$, transitions from below 
unity at lower $p_{\mathrm{T0}}$ values to well above unity at higher $p_{\mathrm{T0}}$. 
Consequently, lower values of $p_{\mathrm{T0}}$ yield a non-flow signal that decreases with increasing $N_{\mathrm{ch}}$, a trend that aligns with the traditional expectation of $1/N_{\mathrm{ch}}$ scaling. For higher values of $p_{\mathrm{T0}}$, the non-flow coefficients eventually increase with $N_{\mathrm{ch}}$. 

\section{Jet production in different multiplicities}

\begin{figure}
\includegraphics[width=0.48\textwidth]{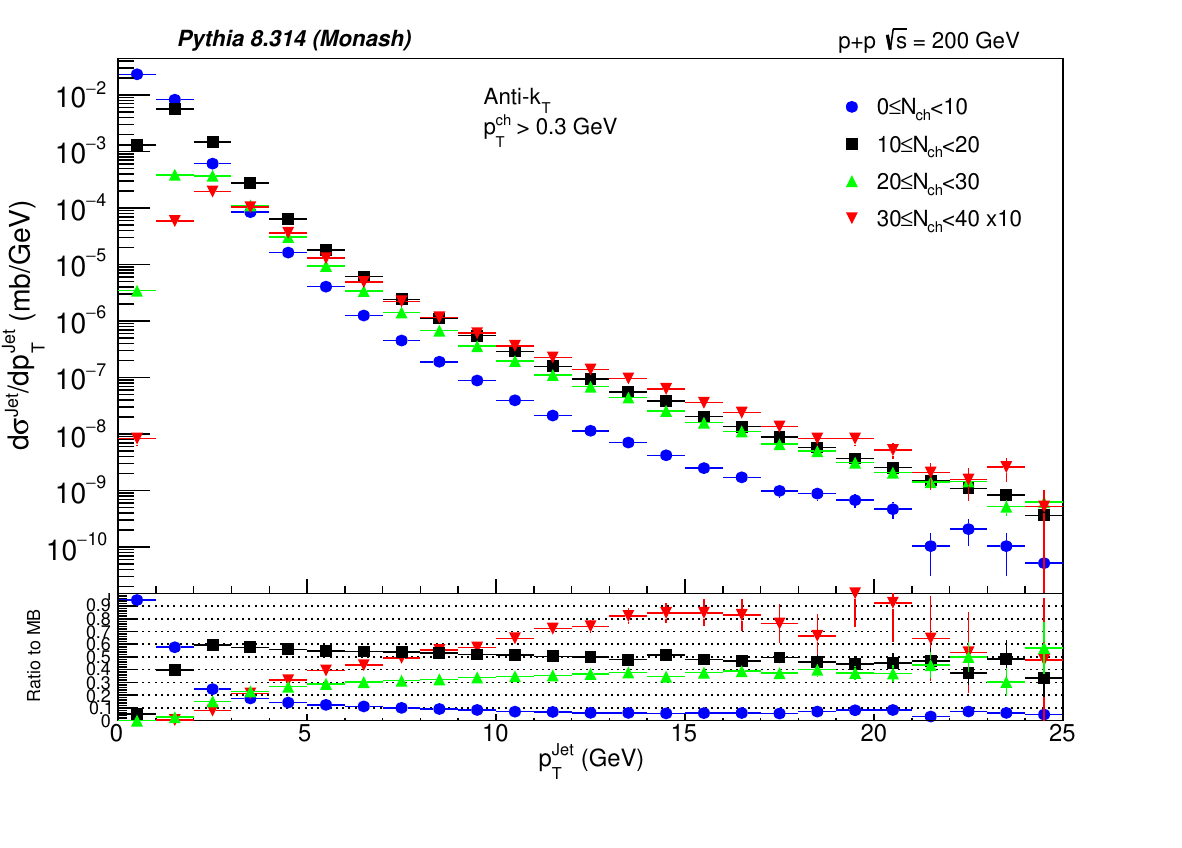}
 \caption{The \pt-differential jet cross-section, measured in different event multiplicity classes.\label{fig:jetpt}}
 \end{figure}

The $p_{\mathrm{T}}\!$-differential jet cross section is studied across different \nch and presented in Fig.~\ref{fig:jetpt}. For each multiplicity interval, the cross section is normalized to the total number of events without multiplicity selection, here referred to as Minimum Bias (MB) events, and the corresponding MB pp luminosity, such that the sum over all multiplicity classes reproduces the total MB $p_{\mathrm{T}}\!$-differential jet cross section. 
The results suggest that at low jet transverse momenta, the shape of the distribution exhibits a strong multiplicity dependence. In this low-\pt region the jet reconstruction is potentially less robust, due to soft particle contamination. Nonetheless, the observed differences reflect genuine modifications to particle fragmentation, which directly modify the structure of non-flow correlations. 

For $p_{\mathrm{T}}^{\mathrm{Jet}} \gtrsim 10$~GeV, where jet reconstruction is expected to be robust, the multiplicity dependence becomes weak for $\nch<30$. However, in the highest multiplicity bin, a significant dependence remains throughout the observed $\pt^{\mathrm{Jet}}$ range. 
Although events containing high-$p_{\mathrm{T}}$ jets are rare, the probability of 
producing such an event varies dramatically with multiplicity. For example, integrating the yield for $p_{\mathrm{T}}^{\mathrm{Jet}} > 10$~GeV relative to the 
total yield shows that the fraction of events containing such a jet in the 
highest multiplicity class ($30 \le N_{\mathrm{ch}} < 40$) is roughly two orders of 
magnitude higher than in the minimum-bias sample and about three orders of magnitude 
higher than in the lowest multiplicity class ($0 \le N_{\mathrm{ch}} < 10$), demonstrating 
a substantial selection bias toward hard processes in high-multiplicity events.

 \begin{figure}
\includegraphics[width=0.48\textwidth]{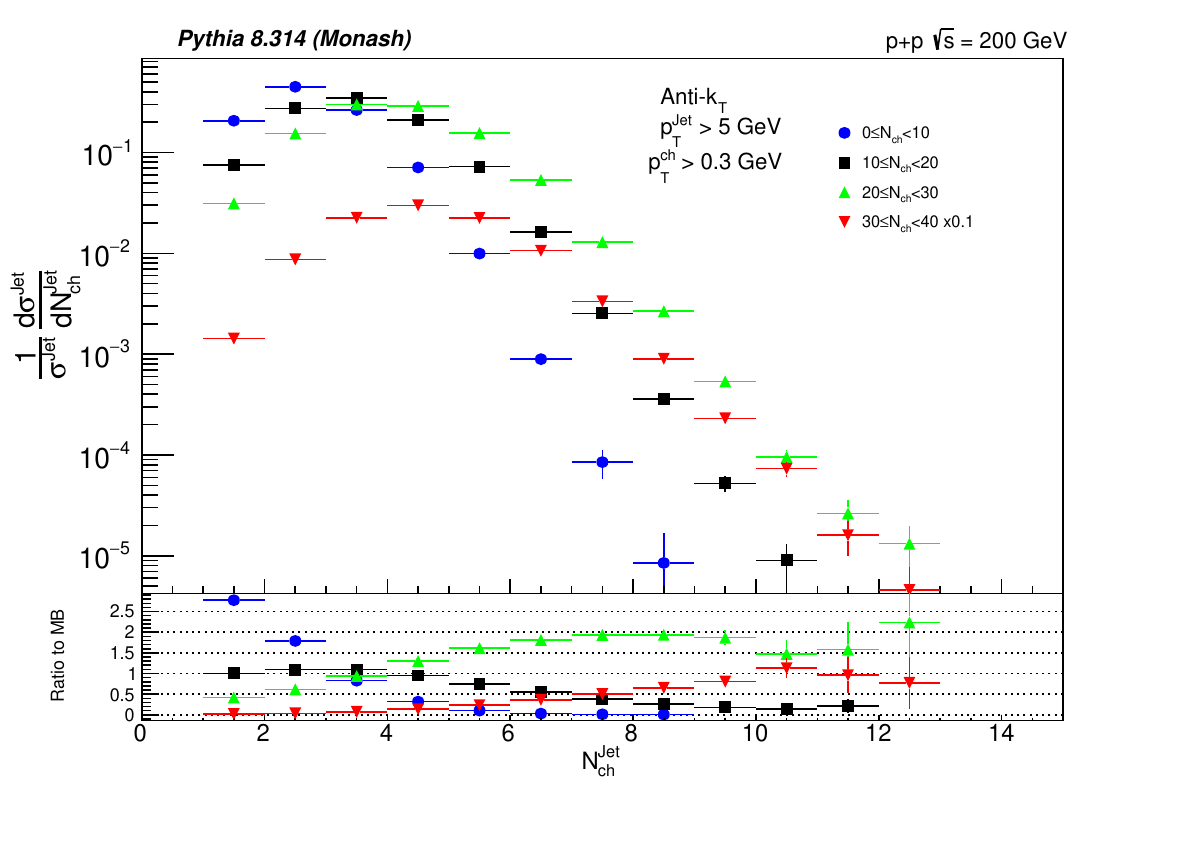}
 \caption{Jet multiplicity distribution (for $p_{\mathrm{T}}^{\mathrm{Jet}} > 5$~GeV) normalized to the total cross section, shown for different event multiplicity classes.\label{fig:jetmult}}
 \end{figure}

To further quantify the interplay between jet production and event activity, the cross section is also studied as a function of jet multiplicity ($\nch^{\text{Jet}}$), normalized per jet, as shown in Fig.~\ref{fig:jetmult}. In this representation, the observable corresponds to the relative probability of producing a jet with particular number of constituent hadrons. The results show a strong dependence of jet multiplicity (for jets with $p_{\mathrm{T}} > 5$~GeV) on event multiplicity, indicating that the event activity selection introduces a significant bias on the sampled jet population.

These results suggest that the initial hard parton production is only weakly dependent on event multiplicity. 
That implies that selecting events based on the number of charged-particles does not strongly modify the rate of hard scatterings at the partonic level, since these processes are set predominantly by the initial parton distribution functions and the hard scattering kinematics rather than by the later soft particle production.
In contrast, the later-stage processes, reflected in the jet fragmentation, are strongly influenced by the imposed multiplicity boundaries. 
Large \nch selections preferentially sample events in which jets fragment more softly, producing a larger number of final-state particles.
Since jet-induced correlations depend on both the number of jets present in the event and their internal structure, 
any selection bias that modifies either of these components will directly propagate into the structure of two-particle correlations and, consequently, the observed non-flow contributions.

\begin{figure}
\includegraphics[width=0.48\textwidth]{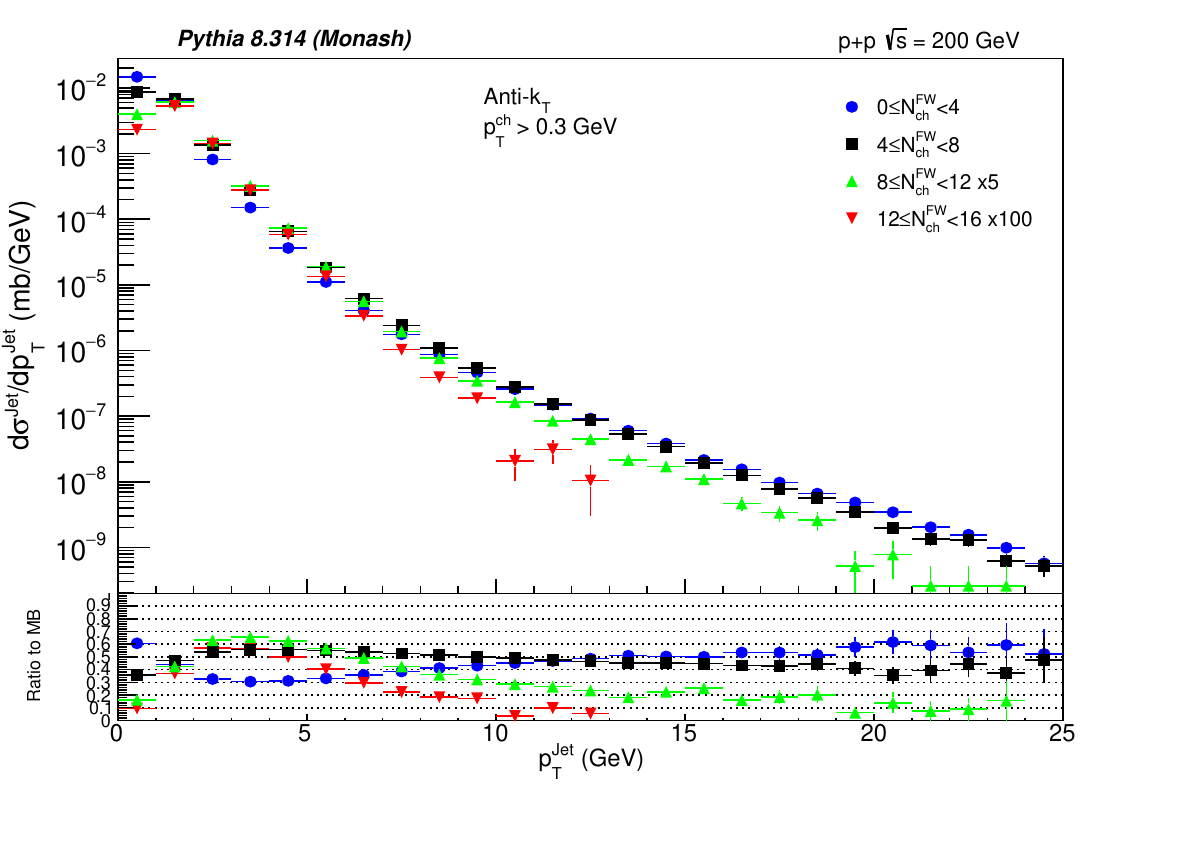}
 \caption{The \pt-differential jet cross-section, measured in different event multiplicity classes, where multiplicity is estimated in forward region $3.5<|\eta|<5$. \label{fig:jetpt_forward}}
 \end{figure}

\begin{figure}
\includegraphics[width=0.48\textwidth]{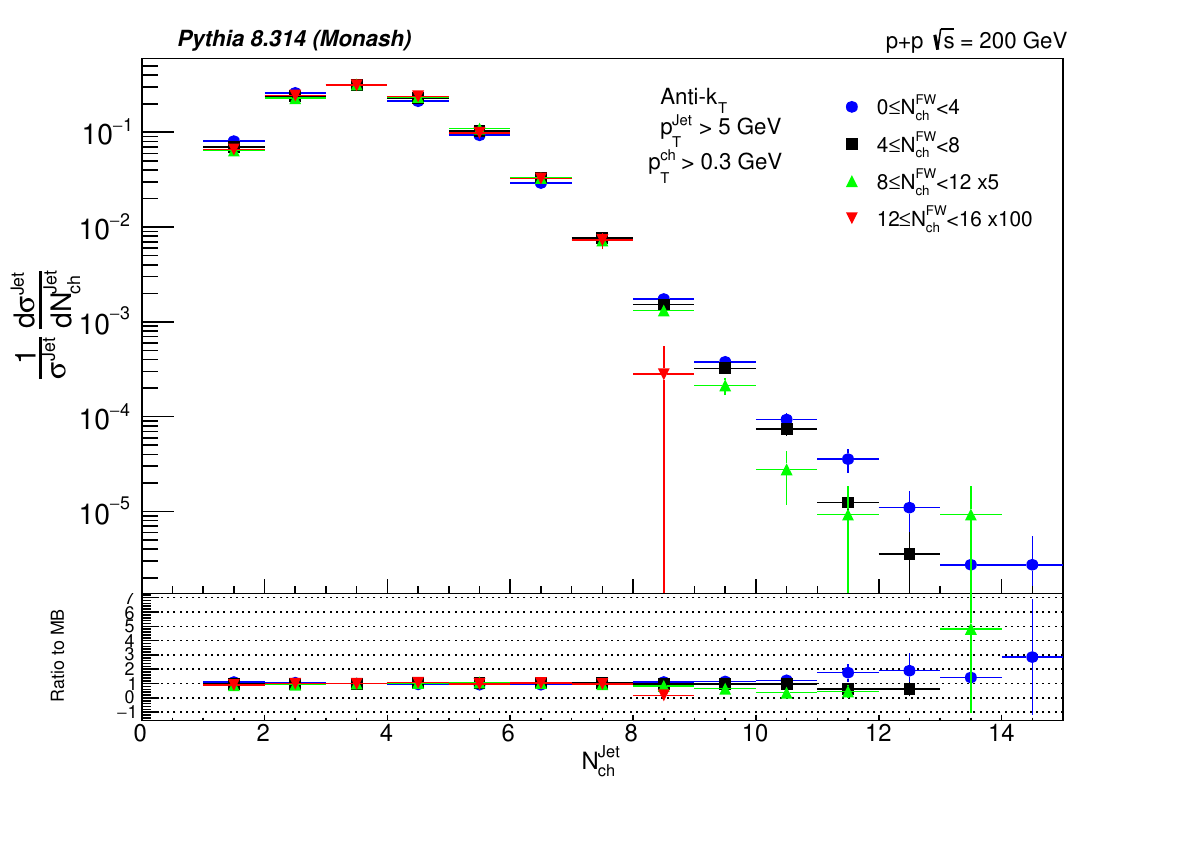}
 \caption{Jet multiplicity distribution (for $p_{\mathrm{T}}^{\mathrm{jet}} > 5$~GeV) normalized to the total cross section, shown for different event multiplicity classes, where multiplicity is estimated in forward region $3.5<|\eta|<5$. \label{fig:jetmult_forward}}
 \end{figure}

To mitigate the potential bias introduced by overlapping the signal region with the event activity estimator, the jet production and particle correlations at midrapidity were studied as a function of \nchfw. The corresponding $p_{\mathrm{T}}$-differential and $\nch^{\text{Jet}}\!$-differential cross sections are shown in Figs.~\ref{fig:jetpt_forward} and \ref{fig:jetmult_forward}, respectively.

In this configuration, the jet multiplicity, as expected, is largely insensitive to the event multiplicity selection, indicating that the direct bias on the reconstructed jet population is significantly reduced. However, the $p_{\mathrm{T}}$-differential jet cross section exhibits a clear dependence on the forward multiplicity. In particular, events with higher particle production in the forward region are less likely to contain high-$p_{\mathrm{T}}$ jets at midrapidity. These correlations between forward particle production and hard scattering kinematics at midrapidity have been reported by STAR Collaboration in pAu collisions~\cite{STAR:2024nwm}. 

\begin{figure}
\includegraphics[width=0.48\textwidth]{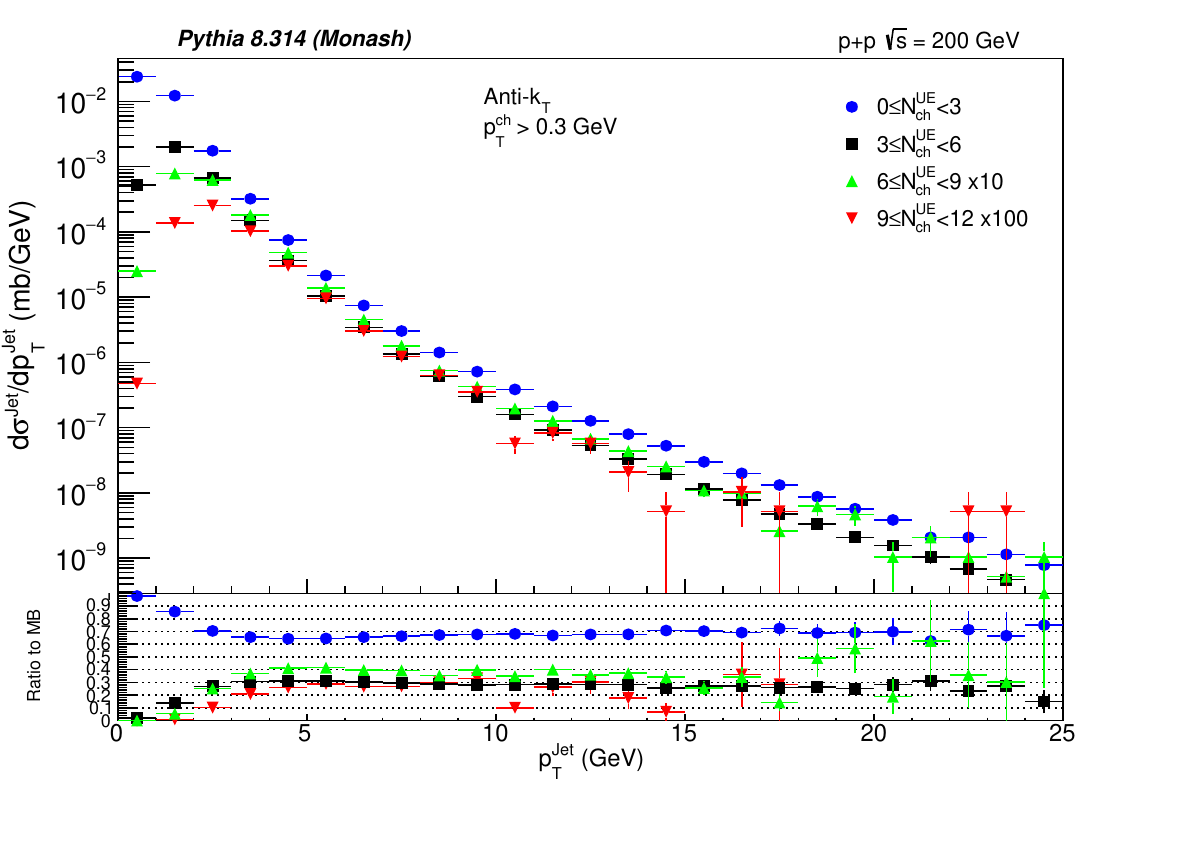}
 \caption{The \pt-differential jet cross-section, measured in different event multiplicity classes, where multiplicity is estimated from the underlying event only.\label{fig:jetpt_ue}}
 \end{figure}

\begin{figure}
\includegraphics[width=0.48\textwidth]{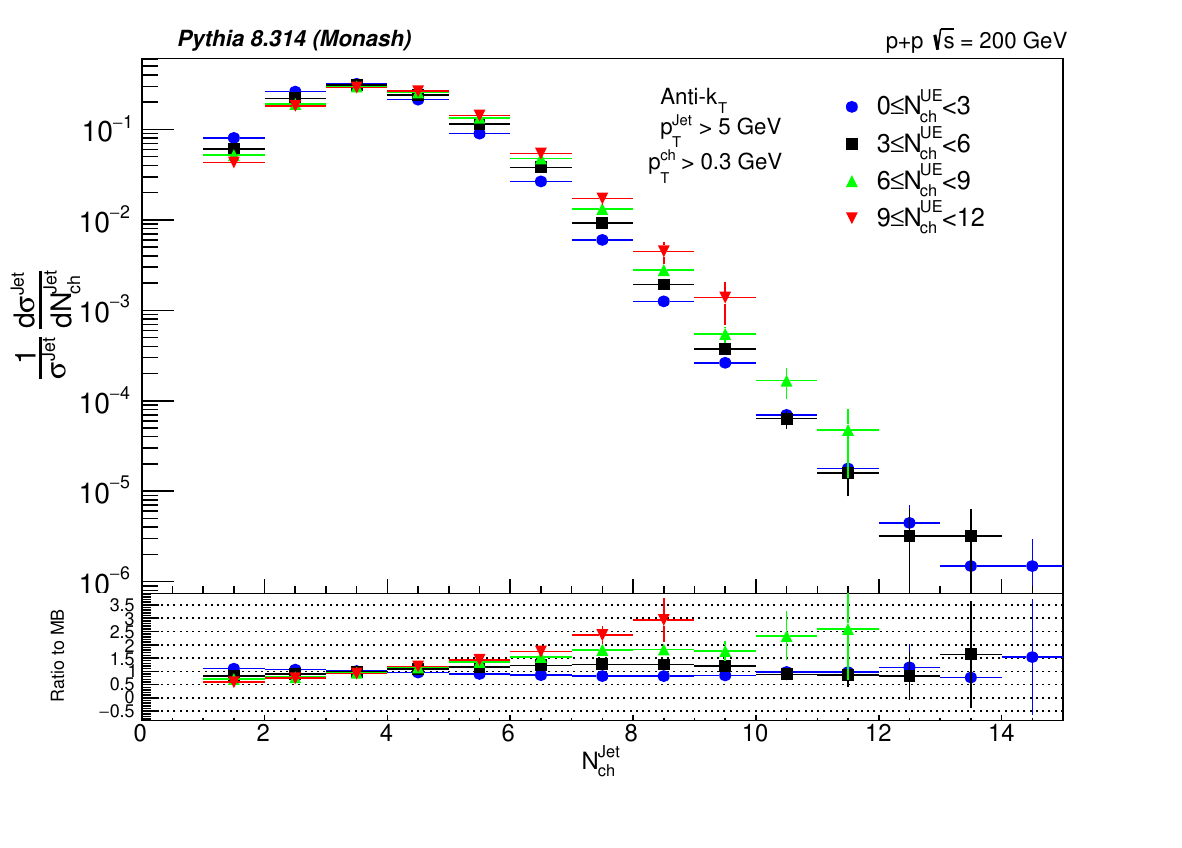}
 \caption{Jet multiplicity distribution (for $p_{\mathrm{T}}^{\mathrm{Jet}} > 5$~GeV) normalized to the total cross section, shown for different event multiplicity classes, where multiplicity is estimated from the underlying event only. \label{fig:jetmult_ue}}
 \end{figure}

\begin{figure}
 \includegraphics[width=0.48\textwidth]{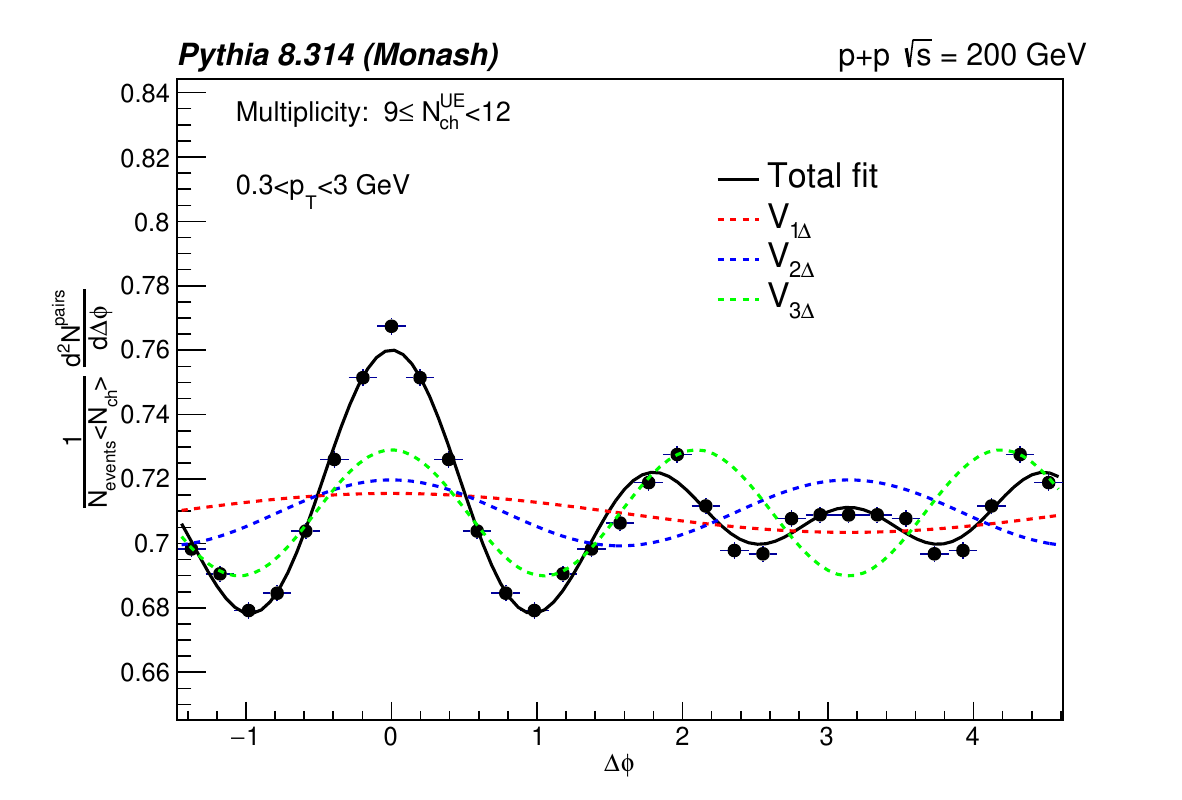}
 \caption{Two-particle azimuthal correlation function as a function of $\dphi$, for $9\leq\nch^{UE}<12$, where $\nch^{UE}$ is number of charged particles in the underlying event. The lines represent the results of Fourier fit.\label{fig:Fourier_UE4main}}
 \end{figure}

Figures~\ref{fig:jetpt_ue} and \ref{fig:jetmult_ue} show $\pt^{\text{Jet}}$ and $\nch^{\text{Jet}}$
as a function of \nchue. In this case, the $p_{\mathrm{T}}$-differential jet cross section exhibits almost no dependence on multiplicity for jet $\pt>3$~GeV, while the jets with larger number of constituents are still more likely to occur in events with high \nchue, but significantly reduced with respect to the \nch definition. At the same time, selecting events with high underlying-event multiplicity introduces a bias on the overall event topology. Requiring a large number of particles within a limited, but symmetrical, azimuthal acceptance selects events in which particle production is already distributed anisotropically in azimuth, as illustrated schematically in Fig.~\ref{fig:cartoonAll}. Figure~\ref{fig:Fourier_UE4main} shows that this selection modifies the shape of the two-particle correlation function, limiting its usefulness for using two particle correlation function to study collective effects in small systems. 

\section{Discussion}

The results presented in this work demonstrate that multiplicity selection is correlated with, and hence directly biases, jet production in small systems at RHIC energies. This bias manifests itself both in the reconstructed jet multiplicity and in the jet $p_{\mathrm{T}}$ spectrum.
This effect is not trivially removed by separating the phase space used for multiplicity estimation from that used for jet reconstruction. Although such separation reduces certain direct biases, residual correlations persist, 
indicating that the interplay between event activity and jet production is more complex than can be accounted for by simple phase-space considerations. As a consequence, jet-induced correlations exhibit a nontrivial dependence on multiplicity that cannot be captured by a universal scaling assumption at RHIC energies using the PYTHIA Monte Carlo event generator.

These observations are established for simulated pp collisions at RHIC energies, significantly below 
than the collision energies typically studied at the LHC. 
Hence, in the context of the used models and the studied kinematics, the effect can be understood in terms of energy conservation and the limited phase space available for particle production~\cite{Bierlich:2022pfr,JETSCAPE:2024dgu}. At lower collision energies, the total energy available in the event is more constrained, such that a significant concentration of energy in one region, a hard jet production or enhanced particle emission at forward rapidity, necessarily reduces the energy available for particle production elsewhere. This introduces an intrinsic anticorrelation between different regions of phase space and leads to a direct coupling between multiplicity selection and the observed properties of jets and correlations. 

The choice of the underlying-event region, with respect to the leading jet axis, as a multiplicity estimator leads to weaker correlations with jet production compared to estimators defined in the same or forward phase space. Despite the evident correlation between the underlying event and the nominal multiplicity, selecting events by the \nchue  biases the overall topology of the event. In particular, high underlying-event multiplicity preferentially selects events with nontrivial azimuthal structure, which can artificially induce large anisotropies in the two-particle correlation function. As a result, the extracted correlation signals are no longer easily interpretable in the context of collective effects, as the selection itself imprints structures that are not directly related to the underlying physics of interest.

Furthermore, systematically varying the infrared regularization cutoff parameter $p_{\mathrm{T0}}$ allows one to probe non-flow behavior across different relative balances of hard and soft particle production. In particular, pushing into regimes where soft emissions are heavily suppressed, the only remaining mechanism capable of producing very high multiplicities is the fragmentation of exceptionally energetic jets. The relative evolution of the Fourier coefficients \vndelta demonstrates a pronounced sensitivity to this hard-to-soft interplay. This observation supports the interpretation that the breakdown of universal non-flow scaling is due to a selection bias where a given multiplicity class favors particular jet-production patterns.

The presented study is based on PYTHIA modeling of pp collisions at $\sqrt{s} = 200$~GeV and it is possible that in experimental data, particularly in other collision systems, non-flow effects may exhibit different behavior than observed in the presented simulation study. Nevertheless, our study suggests a practical, data-driven strategy for assessing potential biases without entirely relying on scaling assumptions. 
While the absence of strong correlations between event multiplicity and jet-related observables does not, by itself, guarantee a reliable non-flow subtraction, any significant multiplicity dependence of these quantities would indicate that the event selection introduces a bias in jet production and, consequently, in the observed correlation structure. Such studies therefore provide an essential cross-check, offering a fully data-driven way to assess the applicability of non-flow subtraction procedures, whether back-to-back correlations are obtained from low-multiplicity events or from theoretical calculations.

In a broader context, the interplay between hard and soft particle production is particularly important for studying QCD. At RHIC energies, where the total available event energy is significantly more constrained than at the LHC, hard processes can have a substantially larger impact on the global event properties. While, on average, the underlying event can often be treated as a reasonable proxy for the overall soft particle production, selecting rare event classes, like high multiplicity events in this study, can introduce strong biases. At the same time, defining multiplicity estimators in pseudorapidity regions separated from the particles of interest may introduce significant anti-correlations due to energy conservation. These effects also provide important constraints for Monte Carlo event-generator tuning, particularly for modeling the interplay between multiparton interactions, jet fragmentation, and soft particle production at lower collision energies. Studying soft-hard correlations at RHIC energies therefore provides a qualitatively different environment compared to the LHC. Consequently, methodologies originally developed and validated in the high-energy regime of the LHC must be carefully reexamined before being directly applied at lower collision energies.

\section{Summary}

In this work, non-flow effects are studied using simulations based on the PYTHIA event generator, which does not contain hydro-dynamically collective effects. The Fourier coefficients $\vndelta$ are extracted from two-particle $\dphi$ correlation functions. The multiplicity evolution of the first, second, and third harmonics, relative to their values in the lowest multiplicity bin, is studied and demonstrates that non-flow contributions do not follow a simple scaling with event multiplicity. By investigating the underlying origin of this behavior, it is shown that jet cross sections and fragmentation are significantly biased by the imposed multiplicity selection, leading to a nontrivial dependence of jet-induced correlations on event activity. These jet-production biases persist even when the multiplicity estimator is defined in the forward region or underlying event, indicating that they cannot be easily avoided. While these results highlight the challenges in treating non-flow effects within commonly used subtraction approaches, the analysis also provides a practical, fully data-driven method to assess potential biases. This approach can be applied across different collision systems, enabling experiments to evaluate the reliability of non-flow subtraction procedures without relying on model-dependent assumptions. More generally, the observed correlations between hard and soft particle production constitute an interesting physics problem in their own right. At RHIC energies, the tighter constraints on event energy mean that hard processes exert a more significant influence on global event properties than at the LHC. 

\begin{acknowledgments}
This work was supported by the U.S. Department of Energy Office of Science, Office of Nuclear Physics under Award No. DE-FG02-92ER40713. 
\end{acknowledgments}

\appendix
\section{Correlation Functions for Multiplicity Estimated in Underlying Event \label{sec:AppA}}

 \begin{figure}[ht]
 \includegraphics[width=0.48\textwidth]{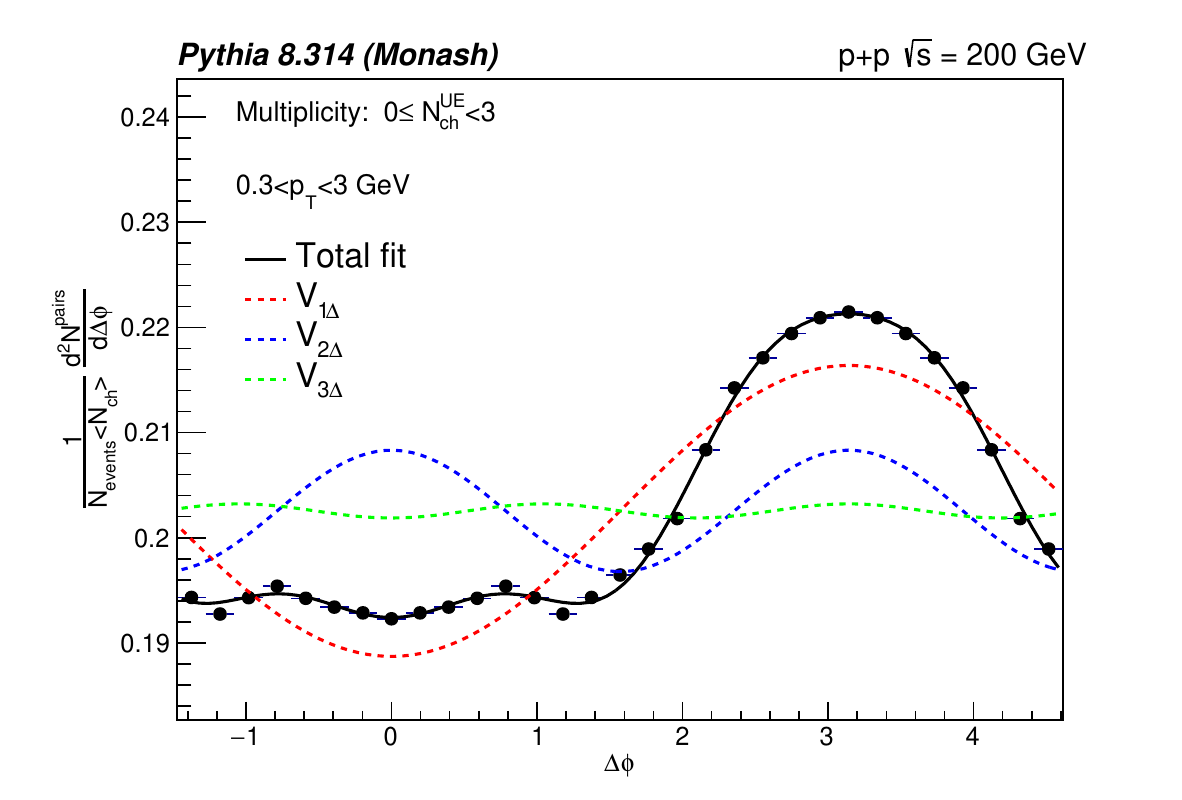}
 \caption{Two-particle azimuthal correlation function as a function of $\dphi$, for $0\leq\nch^{UE}<3$, where $\nch^{UE}$ is number of charged particles in the underlying event. The lines represent the results of Fourier fit.\label{fig:Fourier_UE1}}
 \end{figure}

 \begin{figure}[ht]
 \includegraphics[width=0.48\textwidth]{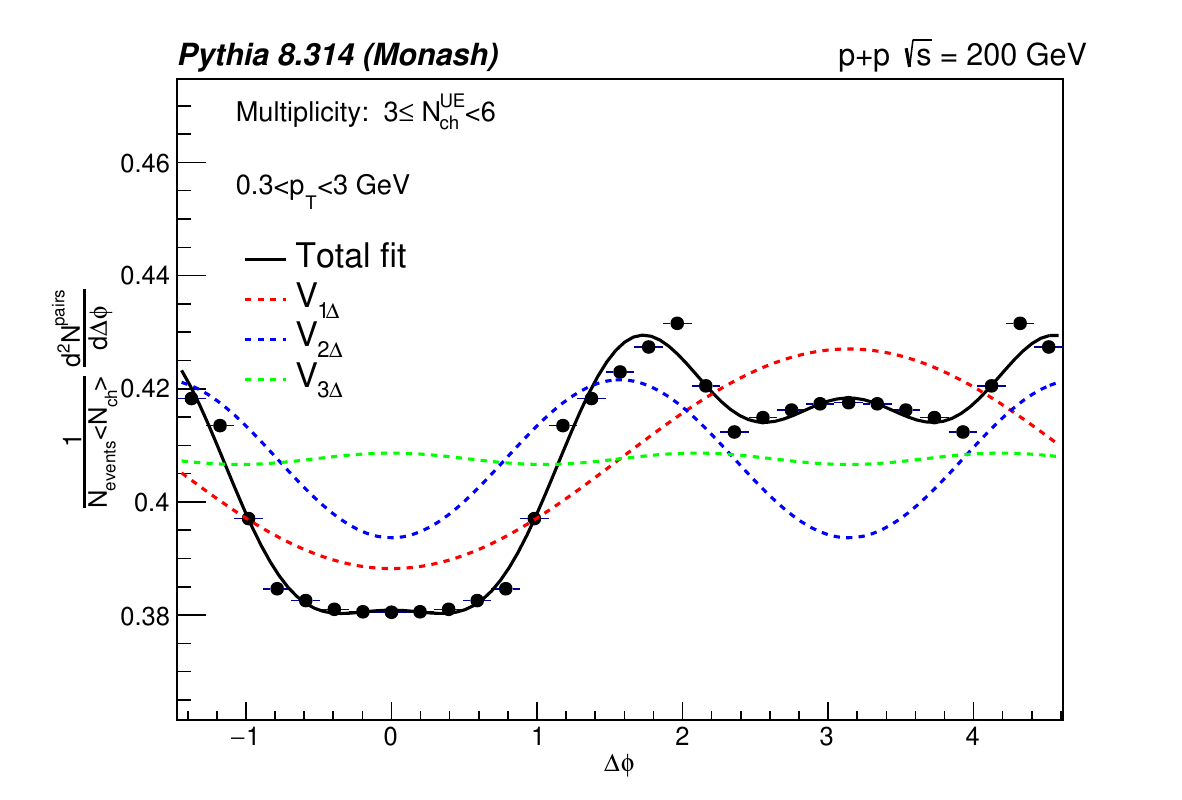}
 \caption{Two-particle azimuthal correlation function as a function of $\dphi$, for $3\leq\nch^{UE}<6$, where $\nch^{UE}$ is number of charged particles in the underlying event. The lines represent the results of Fourier fit.\label{fig:Fourier_UE2}}
 \end{figure}

 \begin{figure}[ht]
 \includegraphics[width=0.48\textwidth]{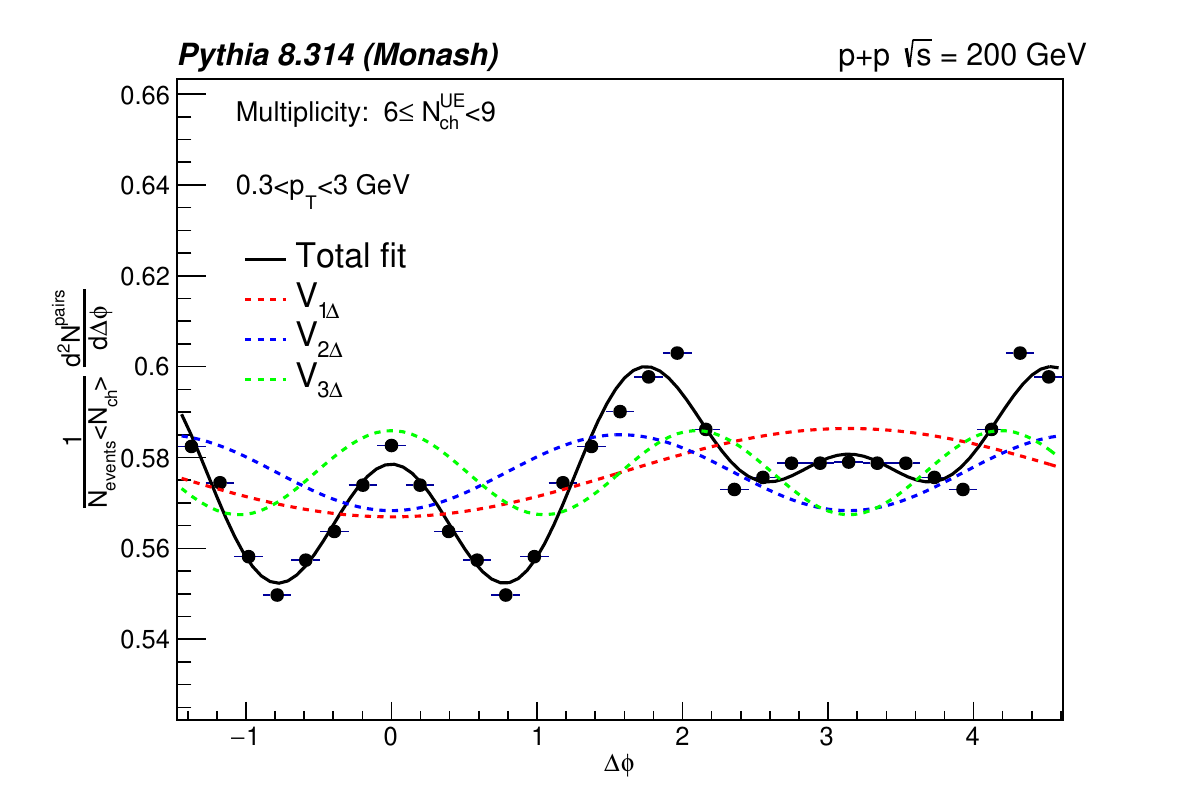}
 \caption{Two-particle azimuthal correlation function as a function of $\dphi$, for $6\leq\nch^{UE}<9$, where $\nch^{UE}$ is number of charged particles in the underlying event. The lines represent the results of Fourier fit.\label{fig:Fourier_UE3}}
 \end{figure}

 \begin{figure}[ht]
 \includegraphics[width=0.48\textwidth]{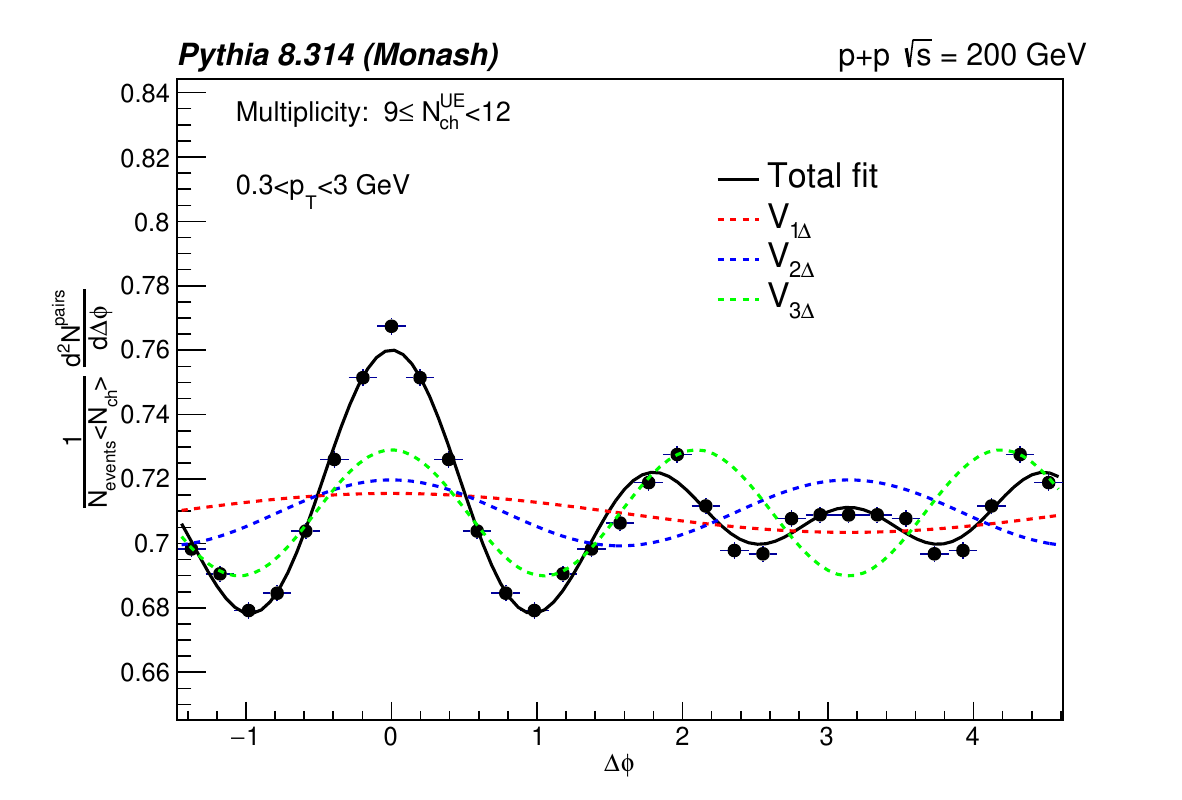}
 \caption{Two-particle azimuthal correlation function as a function of $\dphi$, for $9\leq\nch^{UE}<12$, where $\nch^{UE}$ is number of charged particles in the underlying event. The lines represent the results of Fourier fit.\label{fig:Fourier_UE4}}
 \end{figure}

\section{Dijet Correlations Dependence on Production Mechanisms \label{sec:AppB}}

This appendix provides the charged-particle multiplicity distributions for individual diffractive and non-diffractive processes, along with the multiplicity evolution of the Fourier harmonic coefficients evaluated separately for different types of pp interactions.

 \begin{figure}[ht]
 \includegraphics[width=0.48\textwidth]{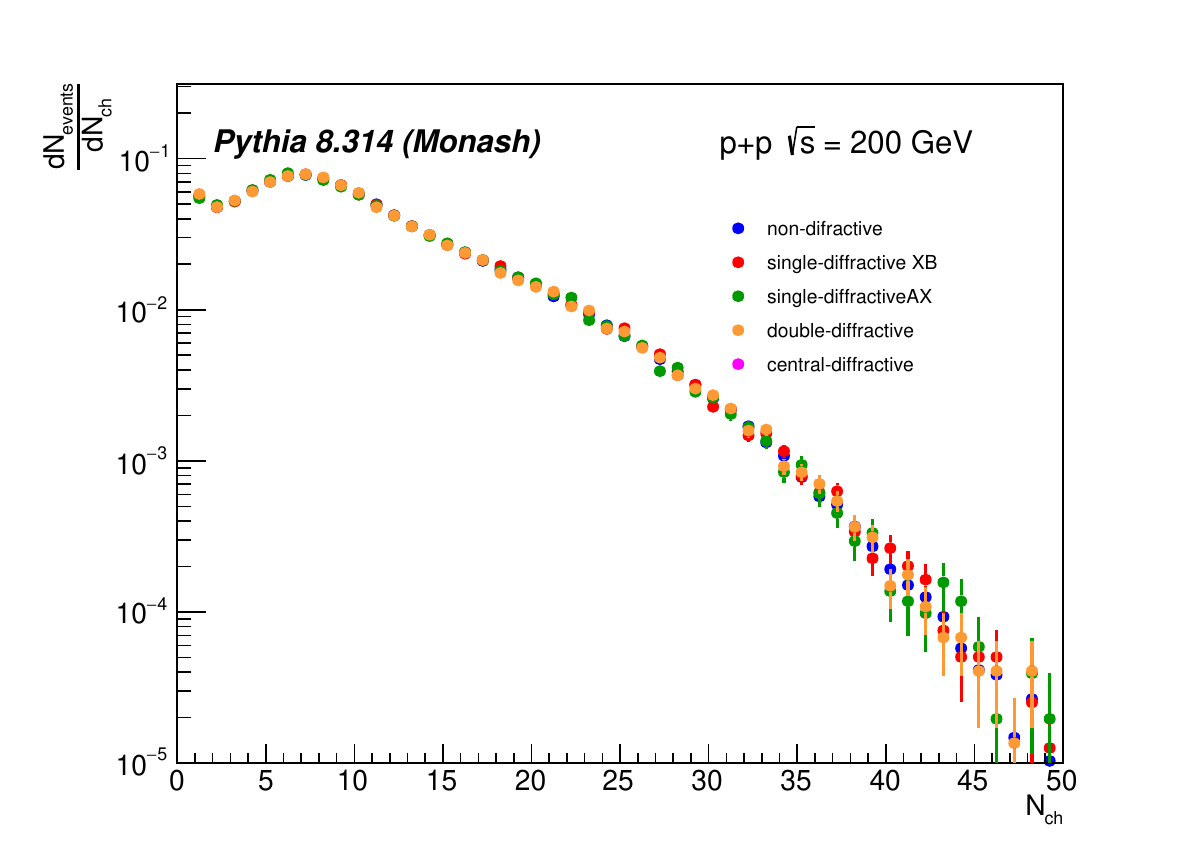}
 \caption{Charged-particle multiplicity distributions decomposed into non-diffractive, single-diffractive ($XB$ and $AX$), double-diffractive, and central-diffractive processes.\label{fig:Mult_appB}}
 \end{figure}

 \begin{figure}[ht]
 \includegraphics[width=0.48\textwidth]{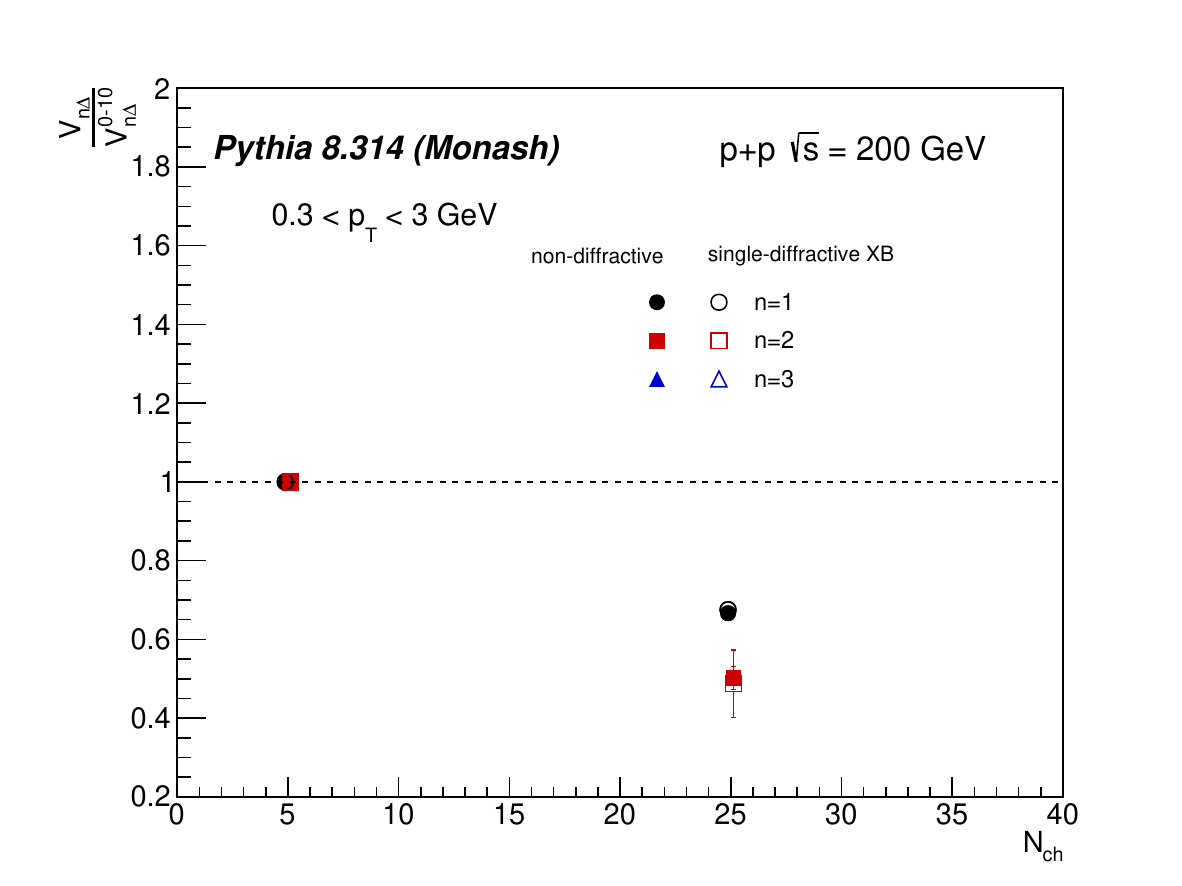}
 \caption{Azimuthal anisotropy coefficients \vonedelta and \vtwodelta, normalized to their value in the lowest multiplicity bin, as a function of event multiplicity for non-diffractive and single-diffractive $XB$ processes.\label{fig:vnratio_appB}}
 \end{figure}

\clearpage

\bibliography{apssamp}

\end{document}